\documentclass[prb,superscriptaddress,showpacs,floatfix]{revtex4-2}
\usepackage{amsfonts}
\usepackage{graphicx,amsmath,amssymb,xspace,epsfig,float,multirow,subfigure,tabularx}
\usepackage{hyperref}
\begin{document}

\title{Modified $GW$ Method in Electronic Systems}
\author{Zhipeng Sun}
\affiliation{School of Physics, Peking University, Beijing 100871, China} 
\affiliation{Collaborative Innovation Center of Quantum Matter, Beijing,
China}
\author{Zhenhao Fan}
\affiliation{School of Physics, Peking University, Beijing 100871, China} 
\affiliation{Collaborative Innovation Center of Quantum Matter, Beijing,
China}
\author{Hui Li}
\affiliation{School of Physics, Peking University, Beijing 100871, China} 
\affiliation{Collaborative Innovation Center of Quantum Matter, Beijing,
China}
\author{Dingping Li}
\email{lidp@pku.edu.cn}
\affiliation{School of Physics, Peking University, Beijing 100871, China} 
\affiliation{Collaborative Innovation Center of Quantum Matter, Beijing,
China}
\author{Baruch Rosenstein}
\email{vortexbar@yahoo.com}
\affiliation{Electrophysics Department, National Yang Ming Chiao Tung University,
Hsinchu 30050, \textit{Taiwan, R. O. C}}

\begin{abstract}
A modified $GW$ approximation to many - body systems is developed. The
approximation has the same computational complexity as the traditional $GW$
approach, but uses a different truncation scheme. This scheme neglects the
high order connected correlation functions. A covariant (preserving the Ward
identities due to the charge conservation) scheme for the two - body
correlators is employed, which holds the relation between the charge
correlator and the charge susceptibility. The method is tested on the two -
dimensional one - band Hubbard model. The results are compared with exact
diagonalization, the $GW$ approximation, the fluctuation - exchange (FLEX)
theory and determinantal Monte Carlo (MC) approach. The comparison for the
(one - body) Green's function demonstrates that it is more precise in strong
- coupling regime (especially away from half - filling) than the $GW$ and
FLEX approximations, which have a similar complexity. More importantly, this
method indicates a Mott - Hubbard gap as the Hubbard $U$ increases, whereas the $GW$
and FLEX methods fail. Besides, the charge correlator obtained from the
covariant scheme not only holds the consistency of the static charge
susceptibility, but also makes a significant improvement over the RPA
calculations.
\end{abstract}

\maketitle

\section{Introduction}

Understanding of physics of the strongly correlated electronic systems has
been a challenge in condensed matter theory for many decades. These systems 
are hosts of 
distinct phenomena such as the Mott insulator [%
\onlinecite{imada_metal-insulator_1998}], quantum magnetism [%
\onlinecite{auerbach_interacting_2012}], pseudogap [%
\onlinecite{timusk_pseudogap_1999}], strange metal [%
\onlinecite{lee_doping_2006}] and $d$-wave high-temperature
superconductivity [\onlinecite{dagotto_correlated_1994}, %
\onlinecite{scalapino_common_2012}], all of which cannot be explained within
the framework of the traditional renormalized weak coupling expansion. Above
the atomic level (described by the density functional approximation), the
main features of these systems are typically captured sufficiently well by
the lattice effective Hamiltonian with (quasi) local Coulomb repulsion. Up
to now, numerous non - perturbative numerical and analytic approaches have
been developed to tackle these seemingly simple models, such as the (one or
multi - band) Hubbard model [\onlinecite{hubbard_electron_1963}].

Numerical non - perturbative methods include the density matrix
renormalization group (DMRG) [\onlinecite{schollwock_density-matrix_2005}],
determinantal quantum Monte Carlo (MC) simulation [%
\onlinecite{becca_quantum_2017}] and dynamic mean-field theory (DMFT) [%
\onlinecite{georges_dynamical_1996}, \onlinecite{kotliar_electronic_2006}].
They can produce reliable results in certain cases, but have limitations in
the cases of interest, for example, at very low temperature or deviations
from half filling (doping). DMRG is reliable mostly in one - dimensional
case, while the determinantal MC encounters a severe fermionic sign problem
and thus fails at low temperature and significant doping. DMFT although
successful at intermediate coupling generally misses nonlocal fluctuations.
A lot of effort was made to remedy this by the extensions of a more
elaborate scheme [\onlinecite{rohringer_diagrammatic_2018}, %
\onlinecite{schaefer_tracking_2021}].

Analytic non - perturbative methods evolved from simple mean field methods [%
\onlinecite{auerbach_interacting_2012}] like variations of Hartree -
Fock(HF), to more sophisticated field theoretical methods. Generally, a
closed set of (quite complicated) equations of the correlators and the
vertex functions is constructed and subsequently solved numerically. Most
used approximations are based on the Baym-Kadanoff formalism [%
\onlinecite{baym_conservation_1961}, \onlinecite{baym_self-consistent_1962}%
], the Hedin's equations [\onlinecite{hedin_new_1965}], the diagrammatic
analysis [\onlinecite{dedominicis_stationary_1964-1}]. Others are based on
particular truncations of Dyson - Schwinger equations [%
\onlinecite{kadanoff_theory_1961, chen_bcs-bec_2005,
rosenstein_covariant_2018}].

By their complexity the analytic methods can be broadly classified into two
classes. In the simpler class, one identifies a function (or functions) of
just \textit{one} energy-momentum variable as the relevant \textquotedblleft
degrees of freedom\textquotedblright . Examples include the electronic
Green's function $G(\omega ,k)$, the screened dynamical potential $W(\omega
,k)$ and the charge and spin susceptibilities $\chi (\omega ,k)$. Beyond the
HF, two popular approximations of this class are the $GW$ approximation [%
\onlinecite{hedin_new_1965}, \onlinecite{aryasetiawan_gw_1998}], involving $%
G $ and $W$, and the fluctuation - exchange (FLEX) theory [%
\onlinecite{bickers_conserving_1989}], involving $G$ and $\chi$'s. More
complicated schemes such as the parquet approximation [%
\onlinecite{dedominicis_stationary_1964}, %
\onlinecite{bickers_conserving_1991}] and covariant quartic approximation [%
\onlinecite{fan_covariant_2020}], in addition to the one - momentum
functions, unfortunately have to consider \textit{multiple} - momenta - 
dependent 
quantities, such as the two - body vertex functions and the high - order
correlators.

To describe realistic correlated materials, the complicated schemes are 
often 
not feasible yet due to their large computational complexity, and thus the
simpler class is more favored. However, the current $GW$ and FLEX
approximation produce less accurate data compared with experimental [%
\onlinecite{kutepov_one-electron_2017}] or numerically exact results [%
\onlinecite{bickers_conserving_1991}]. Therefore, a simpler yet sufficiently
reliable and precise method is highly sought for.

In this paper, one such method, a modification of $GW$ approximation, is
developed. To fully take advantage of the clustering properties of the
connected correlators, the modified $GW$ approximation is 
to truncating 
high order connected correlators on the Dyson - Schwinger equations. The
resulting equations turn out be quite similar to the $GW$ equations, and the
physical meanings are also analogous. As in $GW$ the Coulomb interaction is
renormalized, the screening for long range interaction is included within
the modified $GW$ approximation. In this degree, this method is applicable
to realistic materials.

In a many - body system, the charge conservation leads to a set of the Ward
identities. In an approximation (such as $GW$ or FLEX), the Ward identity
for one - body Green's function is obeyed, whereas the Ward identity for the
two - body correlator (directly obtained from equations after the
approximation) is often violated. Besides, the relation between the charge
correlator and charge susceptibility, $\partial n/\partial \mu =\chi ^{\text{%
ch}}\left( \omega =0,k=0\right) $, is often violated [%
\onlinecite{morita_flex_2002}]. To preserve these identities, the covariant
scheme [%
\onlinecite{kovner_covariant_1989, rosenstein_covariant_1989,
wang_covariant_2017}] is employed in this paper.

The modified $GW$ approximation is tested on the two - dimensional (2D) one
- band Hubbard model in this paper. The results for the density and the
Green's function demonstrate that the modified $GW$ approximation produce
satisfactory results \textit{even in strong - coupling regime}, compared
with exact diagonalization (ED) or determinantal MC approach. 
    The method also indicates a Mott - Hubbard gap as the Hubbard $U$ increases, whereas the $GW$ and FLEX methods fail. 
The results of charge correlator demonstrate a significant improvement over
the RPA scheme, and the charge susceptibility obtained from the covariant
scheme is consistent with an independent calculation $\partial n / \partial \mu$.

This paper is organized as follows. In Sec. \ref{sec:hgw} the modified $GW$
approximation is presented for the fermionic (one - body) Green's function.
Next in Sec. \ref{sec:chgw} the covariant scheme for the two - body
correlators is presented. And then in Sec. \ref{sec:comparison} this method
is tested on the 2D Hubbard model by comparing the (one - body) Green's
function (and the density), and the (two - body) charge correlator (and the
static charge susceptibility) with other approaches. The conclusions and
dissussions are given in Sec. \ref{sec:conclusion}.

\section{Modified $GW$ approximation for fermionic Green's function\label%
{sec:hgw}}

In this section, basic equations and assumptions of the modified $GW$
approximation are presented. The general density - density type interacting
fermionic system at finite temperature is considered. Two exact equations
involving the connected correlators are derived. The approximation is
motivated by the clustering properties of the connected correlators. 

\subsection{Two exact equations for correlators}

The Matsubara action for a density - density type interacting fermionic
system at finite temperature has the form 
\begin{align}
S\left[ \psi ,\psi ^{\ast }\right] & =-\int d\left( 12\right) \
T\left(1,2\right) \psi ^{\ast }\left( 1\right) \psi \left( 2\right)  \notag
\\
& \quad +\frac{1}{2}\int d\left( 12\right) \ V\left( 1,2\right) \rho
\left(1\right) \rho \left( 2\right) \text{,}  \label{eq:general-action}
\end{align}%
where $\psi ^{\ast }$, $\psi $ are Grassmannian fields and $\rho
\left(1\right) \equiv \psi ^{\ast }\left( 1\right) \psi \left( 1\right) $ is
the density (composite operator). The label $\left( 1\right) \equiv \left(
\sigma _{1},x_{1},\tau_{1}\right) $ represents a generalized coordinate,
containing the spin projection $\sigma _{1}$, the space coordinate $x_{1}$,
and the Matsubara time $0<\tau _{1}<\beta $, with $\beta $ being the inverse
temperature. The condensed notation $\int d\left( 1\right) $ stands for the
integral or summation over all the values of a generalized coordinate $%
\left( \sigma_{1},x_{1},\tau _{1}\right) $. \ The bi - local functions $T$
and $V$ are the hopping strength and the interaction (dynamical)
\textquotedblleft potential\textquotedblright . Generalization to several
fermionic species or type of interactions (spin, current) is straightforward.

Consider the perturbation of the system by an external bosonic source $\phi
\left( 1\right) $ (local spin selective chemical potential) coupled to the
density : 
\begin{equation}
S\left[ \psi ,\psi ^{\ast };\phi \right] =S\left[ \psi ,\psi ^{\ast }\right]
-\int d\left( 1\right) \ \phi \left( 1\right) \rho \left( 1\right) \text{.}
\label{eq:perturbed_action}
\end{equation}
Note that unlike  in [\onlinecite{fan_covariant_2020}],  the source is coupled to \ a
quantity quadratic in the fermionic fields. Using the grand partition
function, 
\begin{equation}
Z\left[ \phi \right] =\int \mathcal{D}\left[ \psi ,\psi ^{\ast }\right] \ 
\text{e}^{-S\left[ \psi ,\psi ^{\ast };\phi \right] }\text{,}
\label{eq:partition_function_for_perturbed_system}
\end{equation}%
the (one - body) Green's function $G$ is given by: 
\begin{equation}
G\left( 1,2\right) \equiv \left\langle \psi ^{\ast }\left( 2\right) \psi
\left( 1\right) \right\rangle =\frac{1}{Z\left[ \phi \right] }\int \mathcal{D%
}\left[ \psi ,\psi ^{\ast }\right] \ \psi ^{\ast }\left( 2\right) \psi
\left( 1\right) \text{e}^{-S\left[ \psi ,\psi ^{\ast };\phi \right] }\text{.}
\label{eq:definition-of-green-function}
\end{equation}%
Here $\int \mathcal{D}\left[ \psi ,\psi ^{\ast }\right] $ is the
(Grassmannian) functional path integral measure.

The Green's function $G\left( 1,2\right) $ and its functional derivative $%
\delta G\left( 1,2\right) /\delta \phi \left( 3\right)$ are related through
the following equation of motion (see Appendix \ref{sec:derivation-eom}, for
derivation): 
\begin{equation}
\delta \left( 1,2\right) = \int d\left( 3\right) \ H^{-1}\left( 1,3\right)
G\left( 3,2\right) -\int d\left( 3\right) \ V\left( 1,3\right) \frac{\delta
G\left( 1,2\right) }{\delta \phi \left( 3\right) }\text{.}
\label{eq:equation-of-motion}
\end{equation}%
Here $\delta \left( 1,2\right) $ is the Dirac/Kronecker delta function and
the \textquotedblleft Hartree\textquotedblright \ propagator $H$ is defined
by 
\begin{equation}
H^{-1}\left( 1,2\right) \equiv T\left( 1,2\right) + \delta \left( 1,2\right)
v\left( 1\right) \text{.}  \label{eq:definition_of_h}
\end{equation}%
Here the density weighted interaction potential $v$ is: 
\begin{equation}
v\left( 1\right) \equiv \phi \left( 1\right) -\int d\left( 2\right) \
V\left( 1,2\right) \rho \left( 2\right) \text{.}  \label{eq:definition_of_v}
\end{equation}
Note that in the absence of the external source, i.e. $\phi=0$, the quantity 
$H$ is the free Green function with Hartree self - energy absorbed in the
chemical potential. 

Functional derivative of Eq.\eqref{eq:equation-of-motion} with respect to
the source $\phi$ yields, 
\begin{align}
0& =\int d\left( 4\right) \ \frac{\delta H^{-1}\left( 1,4\right) }{\delta
\phi \left( 3\right) }G\left( 4,2\right) +\int d\left( 4\right) \
H^{-1}\left( 1,4\right) \frac{\delta G\left( 4,2\right) }{\delta \phi \left(
3\right) }  \notag \\
& \quad -\int d\left( 4\right) \ V\left( 1,4\right) \frac{\delta ^{2}G\left(
1,2\right) }{\delta \phi \left( 3\right) \delta \phi \left( 4\right) }\text{,%
}  \label{eq:eom-second}
\end{align}%
relating the one - body correlator $G$ and the two - body correlator
correlator $\delta G/\delta \phi $ to the three - body correlator $%
\delta^{2}G/\delta \phi ^{2}$. By successive functional derivatives, one can
obtain a hierarchy of such relations for even higher order correlators.
These are used in the search of successful non - perturbative approximations
by truncating certain terms considered small by a certain qualitative
argument valid for a particular class of systems and values of parameters.
This way a closed set of equations obtained and solved numerically typically
by iterations.

\subsection{Clustering property of connected correlators and the $HGW$
truncation}

The simplest approximation is to truncating $\delta G/\delta \phi $ in the
first equation of motion Eq.\eqref{eq:equation-of-motion}. This yields $G=H$%
, namely the Hartree approximation widely used in condensed matter physics [%
\onlinecite{auerbach_interacting_2012}]. A more complicated (and hopefully
precise, see below) approximation would be truncating the $\delta
^{2}G/\delta \phi ^{2}$ term in Eq.\eqref{eq:eom-second}. Justification for
such a truncation originates from the clustering property, which states that
the connected correlation function is very small as its coordinates are
separated. The quantity 
\begin{equation}
\frac{\delta ^{2}G\left( 1,2\right) }{\delta \phi \left( 3\right) \delta
\phi \left( 4\right) }=\left \langle \psi ^{\ast }\left( 2\right) \psi
\left( 1\right) \rho \left( 3\right) \rho \left( 4\right) \right \rangle _{%
\text{c}}  \label{eq:three-body-connected-correlator}
\end{equation}%
is a connected correlation function, and thus can be omitted in certain
cases. The reliability of this truncation is determined by the inequality 
\begin{equation}
\left \vert \int d\left( 45\right) H\left( 1,5\right) V\left( 5,4\right) 
\frac{\delta ^{2}G\left( 1,2\right) }{\delta \phi \left( 3\right) \delta
\phi \left( 4\right) }\right \vert \ll \left \vert \int d(45)\ H\left(
1,4\right) \frac{\delta H^{-1}\left( 4,5\right) }{\delta \phi \left(
3\right) }G\left( 5,2\right) \right \vert .  \label{eq:inequality-hgw}
\end{equation}%
Then one can approximate Eq.\eqref{eq:eom-second} by 
\begin{equation}
\frac{\delta G\left( 1,2\right) }{\delta \phi \left( 3\right) }=-\int d(45)\
H\left( 1,4\right) \frac{\delta H^{-1}\left( 4,5\right) }{\delta \phi \left(
3\right) }G\left( 5,2\right) .  \label{eq:hgw-approximation}
\end{equation}%
The validity of inequality (\ref{eq:inequality-hgw}) will be indirectly
checked by whether the Green's function obtained within the approximation is
in good agreement with the numerically exact results.

Eqs.(\ref{eq:equation-of-motion}, \ref{eq:hgw-approximation}) form a closed
set, and will yield the $HGW$ equations (for derivation, see Appendix \ref%
{sec:derivation-hgw-equations}): 
\begin{subequations}
\label{eq:hgw_equations}
\begin{align}
G^{-1}\left( 1,2\right) &= H^{-1}\left( 1,2\right) -\Sigma \left( 1,2\right)
,  \label{eq1} \\
\Sigma \left( 1,2\right) &= -H\left( 1,2\right) W\left( 2,1\right) ,
\label{eq2} \\
W^{-1}\left( 1,2\right) &= V^{-1}\left( 1,2\right) -\Pi \left( 1,2\right) ,
\label{eq3} \\
\Pi \left( 1,2\right) &= H\left( 1,2\right) G\left( 2,1\right) .  \label{eq4}
\end{align}
Apparently, these equations resemble those of the $GW$ approximation (see
Ref.[\onlinecite{aryasetiawan_gw_1998}] or appendix \ref{sec:derivation-gw}%
). The equations Eq.(\ref{eq1}) for Green's function $G$ and Eq.(\ref{eq3})
for screened dynamical potential $W$ are the same, whereas the equations Eq.(%
\ref{eq2}) for self energy function $\Sigma $ and Eq.(\ref{eq4}) \ for
polarization function $\Pi $ are different. The $HGW$ approximation is named
due to its the similarity to the $GW$ approximation and Hartree
approximation. Some of the propagators $G$ in $GW$ equations are replaced by
the Hartree propagator $H$ in the $HGW$ equations.

Essentially, the $HGW$ equations and $GW$ equations are based on different
approximation schemes. The $HGW$ equations are derived by the truncation of
high order connected correlators, whereas the $GW$ equations are based on
simplification of Hedin's vertex. The comparison of these two sets of
equations is summarized in Table \ref{tab:comparison}.

\begin{table}[h]
\caption{Comparison between the $HGW$ and $GW$ equations}
\label{tab:comparison}\centering
\begin{tabular}{|p{3cm}<{\centering}|p{2.5cm}<{\centering}|p{2.5cm}<{\centering}|}
\hline
& $HGW$ & $GW$ \\ \hline\hline
Equation for $G$ & \multicolumn{2}{c|}{$G^{-1}=H^{-1}-\Sigma$} \\ \hline
Equation for $\Sigma$ & $\Sigma = -H W$ & $\Sigma = - G W$ \\ \hline
Equation for $W$ & \multicolumn{2}{c|}{$W^{-1}=V^{-1}-\Pi$} \\ \hline
Equation for $\Pi$ & $\Pi = H G$ & $\Pi = G G$ \\ \hline
Approximation & $\delta^{2} G / \delta \phi^{2} = 0$ & $\Lambda = \hat{1}$
\\ \hline
\end{tabular}%
\end{table}
These formulas will be used to calculate the one - body Green's functions
and the particle density in Sec. \ref{sec:comparison}. Now we turn to more
complicated many - body correlators.

\section{Covariant $HGW$ approximation for the two - body correlator \label%
{sec:chgw}}

In this section, the covariant scheme is employed for the two - body
correlators within the $HGW$ approximation. The covariant $HGW$ equations
for the density - density correlators are also derived by functional
derivatives of the $HGW$ equations.

\subsection{Ward identities and covariance}

In a many - body system, the charge conservation leads to a set of the Ward
identities (see Appendix \ref{sec:ward-identity}). In an approximation (such
as $GW$ or FLEX), the Ward identity for one - body Green's function is
obeyed, whereas the Ward identity for the two - body correlator (directly
obtained from equations after the approximation) is often violated. Besides,
the relation between the charge correlator and charge susceptibility, $%
\partial n/\partial \mu =\chi^{\text{ch}} \left(q=0,\omega =0\right)$, is
also often violated. To preserve the consistency in the $HGW$ approximation,
one can define the two - body (connected) correlator as the functional
derivative of Green's function $G$ with respect to the external source $\phi 
$: 
\end{subequations}
\begin{equation}
L^{\text{cov}}\left( 1,2;3\right) = \left. \frac{\delta G\left( 1,2\right) 
} {\delta \phi \left( 3\right) }\right \vert _{\phi=0}.
\label{eq:cov-two-body-correlator}
\end{equation}%
Here $G$ is obtained from the off - shell (nonzero $\phi $) equations. The
superscript \textquotedblleft cov\textquotedblright \ in $L^{\text{cov}}$
denotes for \textquotedblleft covariant\textquotedblright .

As $G$ obeys the Ward identity for all $\phi $'s, the derivative of the Ward
identity is also satisfied: 
\begin{equation}
\int d\left( 2\right) \ \left( T\left( 1,2\right) \frac{\delta G\left(
2,1\right) }{\delta \phi \left( 3\right) }-T\left( 2,1\right) \frac{\delta
G\left( 1,2\right) }{\delta \phi \left( 3\right) }\right) =0.
\label{eq:derivative-first-wti}
\end{equation}%
Letting $\phi =0$ in Eq.(\ref{eq:derivative-first-wti}), one obtains: 
\begin{equation}
\int d\left( 2\right) \ \left( T\left( 1,2\right) L^{\text{cov}}\left(
2,1;3\right) -T\left( 2,1\right) L^{\text{cov}}\left( 1,2;3\right) \right)
=0.  \label{eq:second_wti}
\end{equation}%
Thus one arrives at the conclusion that $L^{\text{cov}}$ defined by Eq.(\ref%
{eq:cov-two-body-correlator}) satisfies the Ward identity for the two - body
correlator. In other words, the covariant scheme automatically preserves all
the charge - conserving laws.

\subsection{Covariant $HGW$ equations for the density - density correlator}

The covariant version of the density - density correlator is defined as, 
\begin{equation}
\chi ^{\text{cov}}\left( 1,2\right) =\left. \frac{\delta \rho \left(
1\right) }{\delta \phi \left( 2\right) }\right \vert _{\phi =0},
\label{eq:chi_cov_definition}
\end{equation}%
with $\rho $ the density obtained from off - shell $HGW$ equations Eq.(\ref%
{eq:hgw_equations}). To compute $\chi ^{\text{cov}}$, one differentiates the 
$HGW$ equations with respect to $\phi $. After calculation, given in
Appendix \ref{sec:derivation-covariant-hgw}, one obtains 
\begin{equation}
\chi ^{\text{cov}}\left( 1,2\right) =\chi _{0}\left( 1,2\right) -\int
d\left( 34\right) \ \chi _{0}\left( 1,3\right) V\left( 3,4\right) \chi ^{%
\text{cov}}\left( 4,2\right) ,  \label{eq:chi-cov-chi-zero}
\end{equation}%
where the covariant version of polarization function $\chi _{0}$ satisfies
the equation: 
\begin{equation}
\chi _{0}\left( 1,2\right) =-\int d\left( 34\right) \ G\left( 1,3\right)
G\left( 4,1\right) \Lambda \left( 3,4;2\right) .
\label{eq:chi-zero-cov-equation}
\end{equation}%
The covariant version of vertex function $\Lambda $ satisfies a set of
linear equations (\ref{eq:derivative-equation-lambda}, \ref%
{eq:derivative-equation-gamma}) given in Appendix \ref%
{sec:derivation-covariant-hgw}.

The calculation procedure for $\chi ^{\text{cov}}$ in the covariant $HGW$
approximation therefore can be summarized as follows. First, one solves the
on - shell ($\phi =0$) $HGW$ equations (\ref{eq:hgw_equations}) to obtain $%
H,G,W$. Second, one solves Eqs.(\ref{eq:derivative-equation-lambda}, \ref%
{eq:derivative-equation-gamma}) to obtain $\Lambda $. Third, one uses Eq.(%
\ref{eq:chi-zero-cov-equation}) to calculate $\chi _{0}$. Finally, one
solves Eq.(\ref{eq:chi-cov-chi-zero}) to obtain $\chi ^{\text{cov}}$. These
equations, Eqs.(\ref{eq:derivative-equation-lambda}, \ref%
{eq:derivative-equation-gamma}, \ref{eq:chi-zero-cov-equation}, \ref%
{eq:chi-cov-chi-zero}), are referred to the covariant $HGW$ equations.

Let us contrast this with frequently used RPA formula for $\chi $, 
\begin{equation}
\chi ^{\text{RPA}}\left( 1,2\right) =\overline{\chi }\left( 1,2\right) -\int
d\left( 34\right) \overline{\chi }\left( 1,3\right) V\left( 3,4\right) \chi
^{\text{RPA}}\left( 4,2\right) ,  \label{eq:chi_rpa_chi_0_rpa}
\end{equation}%
where the Lindhard polarization function $\bar{\chi}$ is given by 
\begin{equation}
\overline{\chi }\left( 1,2\right) =-G\left( 1,2\right) G\left( 2,1\right) ,
\label{eq:solution_chi_0_rpa}
\end{equation}%
with $G\left( 1,2\right) $ approximated within a certain approach (such as
the $GW$). Although the RPA scheme is much simpler than the covariant
scheme, it does not guarantee the Ward identities. Besides, the charge
susceptibility (charge susceptibility) $\partial n/\partial \mu $ is not
consistent with that obtained from the RPA calculation [%
\onlinecite{morita_flex_2002}]. In contrast, the covariant scheme preserves
all these identities.

\section{Comparison with other approximations in the 2D Hubbard model \label%
{sec:comparison}}

In this section the $HGW$ approximation is tested on a (numerically)
solvable model, the 2D one - band Hubbard model. Exact diagonalization is
possible on a relatively small cluster $N\times N$, $N=4$, so we mainly
focus on this system. In many cases we use determinantal Monte Carlo (MC) in
the range of parameters in which it is consistent with the exact
diagonalization (practically not too low temperature and not too large $U$).

The discretized time Matsubara action is employed for numerical
implementation to the 2D Hubbard model. Results of the Green's function (and
the density) within the $HGW$ approximation and the charge correlator (and
charge susceptibility) based on the covariant scheme are presented. At any
stage the $HGW$ method is compared to two other relatively simple analytic
approaches, $GW$ and FLEX (generally all the three approximations are much
better than the Hartree - Fock approximation not shown here).

\subsection{Matsubara action for the 2D Hubbard model}

The Hamiltonian of the 2D Hubbard model is: 
\begin{equation}
\hat{\mathcal{H}}=\sum_{ij}\sum_{\sigma =\uparrow ,\downarrow }t_{ij}\hat{%
\psi}_{i,\sigma }^{\dagger }\hat{\psi}_{j,\sigma }+U\sum_{i}\hat{\psi}%
_{i\uparrow }^{\dagger }\hat{\psi}_{i\uparrow }\hat{\psi}_{i\downarrow
}^{\dagger }\hat{\psi}_{i\downarrow }-\mu \sum_{i,\sigma }\hat{\psi}%
_{i,\sigma }^{\dagger }\hat{\psi}_{i,\sigma }.
\label{eq:hubbard-hamiltonian}
\end{equation}%
Here $\hat{\psi}_{i\sigma }^{\dagger },\hat{\psi}_{i\sigma }$ are the
creation and annihilation operators of electron with spin $\sigma =\uparrow
,\downarrow $ on lattice site $i$. The labels $i$, $j$ denote the
coordinates on the $N\times N$ 2D square lattice with periodic boundary
conditions (and lattice constant setting the unit of length). The hopping
strength $t_{ij}$ equals to $-t$, if sites $i,j$ are nearest neighbors and $%
0 $ otherwise. We set $t=1$ to the unit of energy. Coupling $U$ is the on -
site repulsion, and $\mu $ is the chemical potential.

The discretized time Matsubara action [\onlinecite{negele_quantum_2018}] for
Hamiltonian \eqref{eq:hubbard-hamiltonian} has the form: 
\begin{align}
S_{M}\left[ \psi ,\psi ^{\ast }\right] & =\sum_{l=0}^{M-1}\sum_{\sigma
=\uparrow ,\downarrow }\sum_{i}\ \psi _{i\sigma }^{\ast }\left( \tau
_{l}\right) \left( \psi _{i\sigma }\left( \tau _{l+1}\right) -\psi _{i\sigma
}\left( \tau _{l}\right) \right)  \notag \\
& \quad +\Delta \tau \sum_{l=0}^{M-1}\mathcal{H}\left[ \psi _{i\sigma
}^{\ast }\left( \tau _{l}\right) ,\psi _{i\sigma }\left( \tau _{l}\right) %
\right] .  \label{eq:hubbard-action}
\end{align}%
Here $M$ is the number of Matsubara time slices, so that $\Delta \tau \equiv
\beta /M$ is the time step. The discrete label $l$ takes integral value in $%
\left[ 0,M-1\right] $ and $\tau _{l}\equiv l\Delta \tau $. The functional $%
\mathcal{H}$ is obtained by substituting $\psi _{i\sigma }^{\ast }\left(
\tau _{l}\right) ,\psi _{i\sigma }\left( \tau _{l}\right) $ for $\hat{\psi}%
_{i\sigma }^{\dagger },\hat{\psi}_{i\sigma }$ in Hamiltonian $\hat{\mathcal{H%
}}$ respectively.

Comparing the Matsubara action \eqref{eq:hubbard-action} with general action
(\ref{eq:general-action}), and one obtains the expression for the hopping
matrix $T$: 
\begin{equation}
T\left( 1,2\right) =\Delta \tau \delta _{\sigma _{1}\sigma _{2}}\left( -%
\frac{1}{\Delta \tau }\delta _{i_{1}i_{2}}\left( \delta
_{l_{1},l_{2}-1}-\delta _{l_{1},l_{2}}\right) -t_{i_{1}i_{2}}\delta
_{l_{1}l_{2}}+\mu \delta _{i_{1}i_{2}}\delta _{l_{1}l_{2}}\right) ,
\label{eq:expression_for_t}
\end{equation}%
and expression for the two - body interaction potential $V$: 
\begin{equation}
V\left( 1,2\right) =\Delta \tau U\left( 1-\delta _{\sigma _{1}\sigma
_{2}}\right) \delta _{l_{1}l_{2}}\delta _{i_{1}i_{2}},
\label{eq:expression_for_v}
\end{equation}%
where $\left( 1\right) $ denotes $\left( \sigma _{1},i_{1},\tau _{1}\right) $%
%
, which is a collection of the spin, Matsubara time, and lattice coordinate
indexes 
. The correlators in discretized time Matsubara action are discussed in
Appendix \ref{sec:matsubara-action}.

For a given set of parameters $U,\mu ,T\equiv 1/\beta$ (and $N,M$), one
solves the $HGW$ equations Eq.(\ref{eq:hgw_equations}) to obtain the Green's
functions. The $HGW$ equations in frequency - momentum space is given in
Appendix \ref{sec:derivation-hgw-equation-hubbard}, the covariant $HGW$
equations in frequency - momentum space is given in Appendix \ref%
{sec:derivation-covariant-hgw-hubbard}, and the numerical algorithm and cost
are described in Appendix \ref{sec:algorithm}. We start with the
thermodynamics and then proceed to the Matsubara Green's function and the
charge correlator.

\subsection{Doping dependence of the particle density}

\begin{figure}[htbp]
\includegraphics[width=\linewidth]{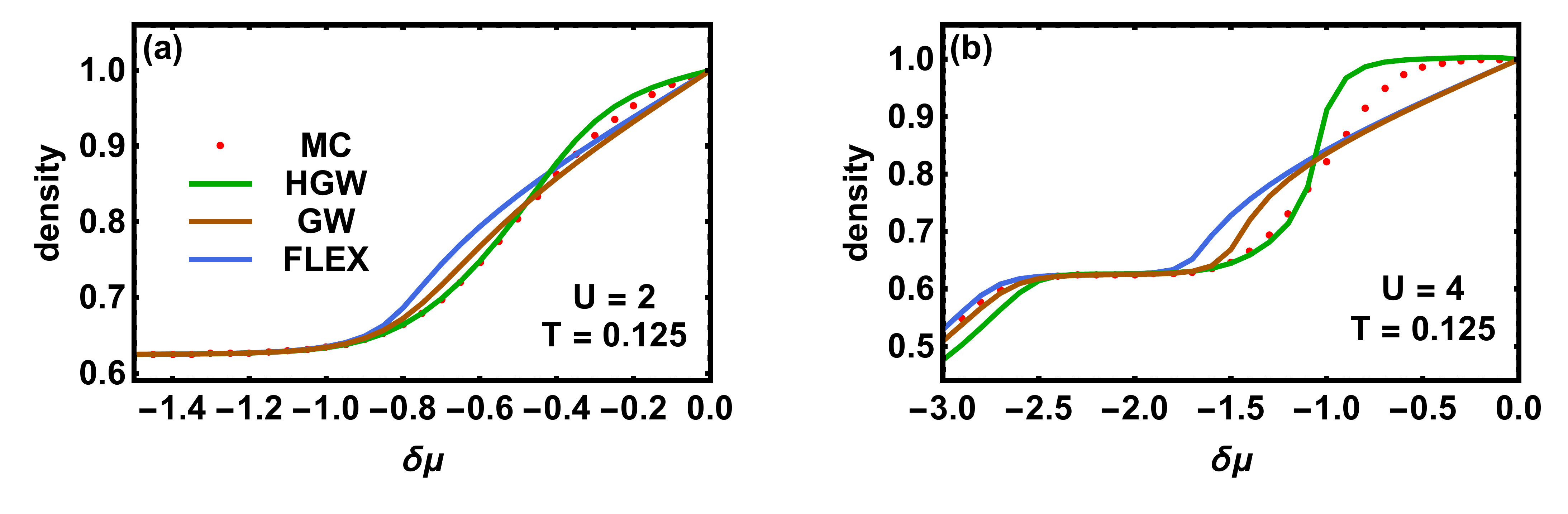}
\caption{The doping $\protect\delta \protect\mu$ dependence of the particle
density at (a) $U=2, T=0.125$ and (b) $U=4, T=0.125$ for the $4\times4$
Hubbard cluster. The red dots denote the results obtained from MC. The
darker green line denotes the results obtained from $HGW$ equations. The
darker orange line denotes the results obtained from $GW$ equations. The
royal blue line denotes the results obtained from FLEX approximation.}
\label{fig: den_vs_doping}
\end{figure}

To study the doping dependence of the particle density we chose $T=0.125$
for $4\times 4$ cluster and two values of the on - site repulsion $U=2$,
representing the weak coupling strength, see Fig.\ref{fig: den_vs_doping}(a)
and $U=4$, representing the intermediate coupling strength, see Fig.\ref%
{fig: den_vs_doping}(b). The results are compared with those obtained from $%
GW$, FLEX and determinantal MC (the ED approach produces the numerically
same results). In Fig.\ref{fig: den_vs_doping}(a), the three curves are all
close to MC result (dots), which means $HGW$, $GW$ and FLEX all produce
satisfactory results of the density at weak coupling regime. In Fig.\ref%
{fig: den_vs_doping}(b), the HGW curve is much closer to MC result than GW
and FLEX when the particle density is larger than $0.6$, which shows $HGW$
is much better than $GW$ and FLEX in the strong antiferromagnetic fluctuation
regime. Besides, MC dots show a plateau resembling for the Mott - Hubbard gap (due
to the strong antiferromagnetic fluctuation) phase near half filling. The HGW
curve exhibits this property, whereas GW and FLEX fails. In this degree, the 
$HGW$ approximation has advantage in capturing the Mott - Hubbard gap over $GW$
and FLEX approximation.

\subsection{Matsubara Green's function}


\subsubsection{Matsubara Green's function at the Matsubara time axis}

\begin{figure}[tbph]
\includegraphics[width=\linewidth]{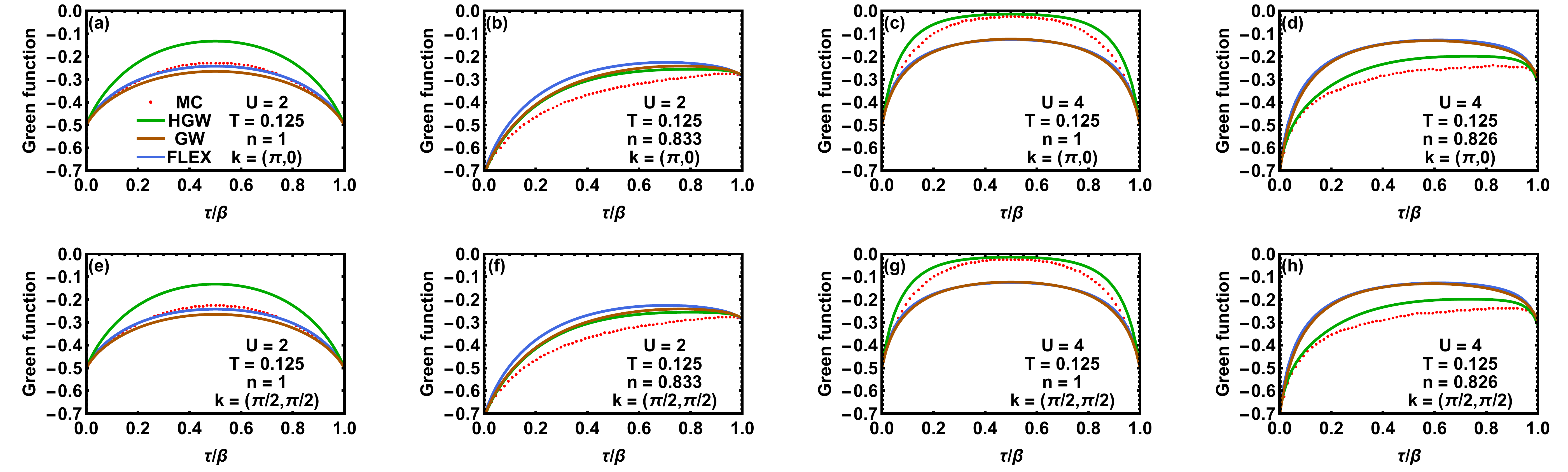}
\caption{Comparison of results of Green's function at Matsubara time axis for $4\times 4$ cluster at
 $T = 0.125$ for different parameters: (a) $U=2,n=1,k=(\pi,0)$%
, (b) $U=2,n=0.833,k=(\pi ,0)$, (c) $U=4,n=1,k=(\pi ,0)$, (d) $U=4,n=0.826,k=(\pi ,0)$,%
(e) $U=2,n=1,k=(\pi/2 ,\pi/2)$, (f) $U=2,n=0.833,k=(\pi/2,\pi/2)$, (g) $U=4,n=1,k=(\pi/2 ,\pi/2)$, (h) $U=4,n=0.826,k=(\pi/2 ,\pi/2)$. The red dots denote the results obtained from MC. The darker green line
denotes the results obtained from the $HGW$ equations. The darker orange line
denotes the results obtained from the $GW$ equations. Royal blue line
denotes the results obtained from the FLEX approximation.}
\label{fig: comparison_green_func_4x4}
\end{figure}

We compare results of Green's function 
at the Matsubara time axis 
at the anti - nodal momentum $k=(\pi ,0)$ and the nodal point $k=(\pi/2,\pi/2)$ (see Fig.\ref{fig:
comparison_green_func_4x4}) for different doping and the coupling strength with 
$T=0.125$ for the $4\times 4$ cluster. At $U=2$ and half - filling (see Fig.\ref{fig:
comparison_green_func_4x4}(a,e)), the GW and FLEX curves are close to the MC data
(dots), whereas the HGW curve is relatively further. At $U=2$ and away from
half - filling (see Fig.\ref{fig: comparison_green_func_4x4}(b,f)), the three
curves are close to each other, but all relatively further away from the MC
result. These results demonstrate that $HGW$ might not be advantageous in
the 
weak coupling regime (particularly at half - filling). 

At a stronger coupling $U=4$, at half - filling (see Fig.\ref{fig:
comparison_green_func_4x4}(c,g)), the HGW curve is much closer to the MC than the
GW and the FLEX curves. As away from half - filling (see Fig.\ref{fig:
comparison_green_func_4x4}(d,h)), the HGW curve is also much closer to the MC data
than the GW and FLEX curves. These results demonstrate that, the $HGW$
approximation has a considerable advantage over $GW$ and FLEX in strong
coupling regime especially away from half - filling.

At half filling, the determinantal MC is applicable to $8\times8$ lattice.
We compare the $HGW$ method in these cases (see Fig.\ref{fig:
comparison_green_func_8x8}). These results also demonstrate that at $U=2$,
the $HGW$ method is worse than the $GW$ and FLEX methods, but is better at $%
U=4$.  
\begin{figure}[tbph]
\includegraphics[width=\linewidth]{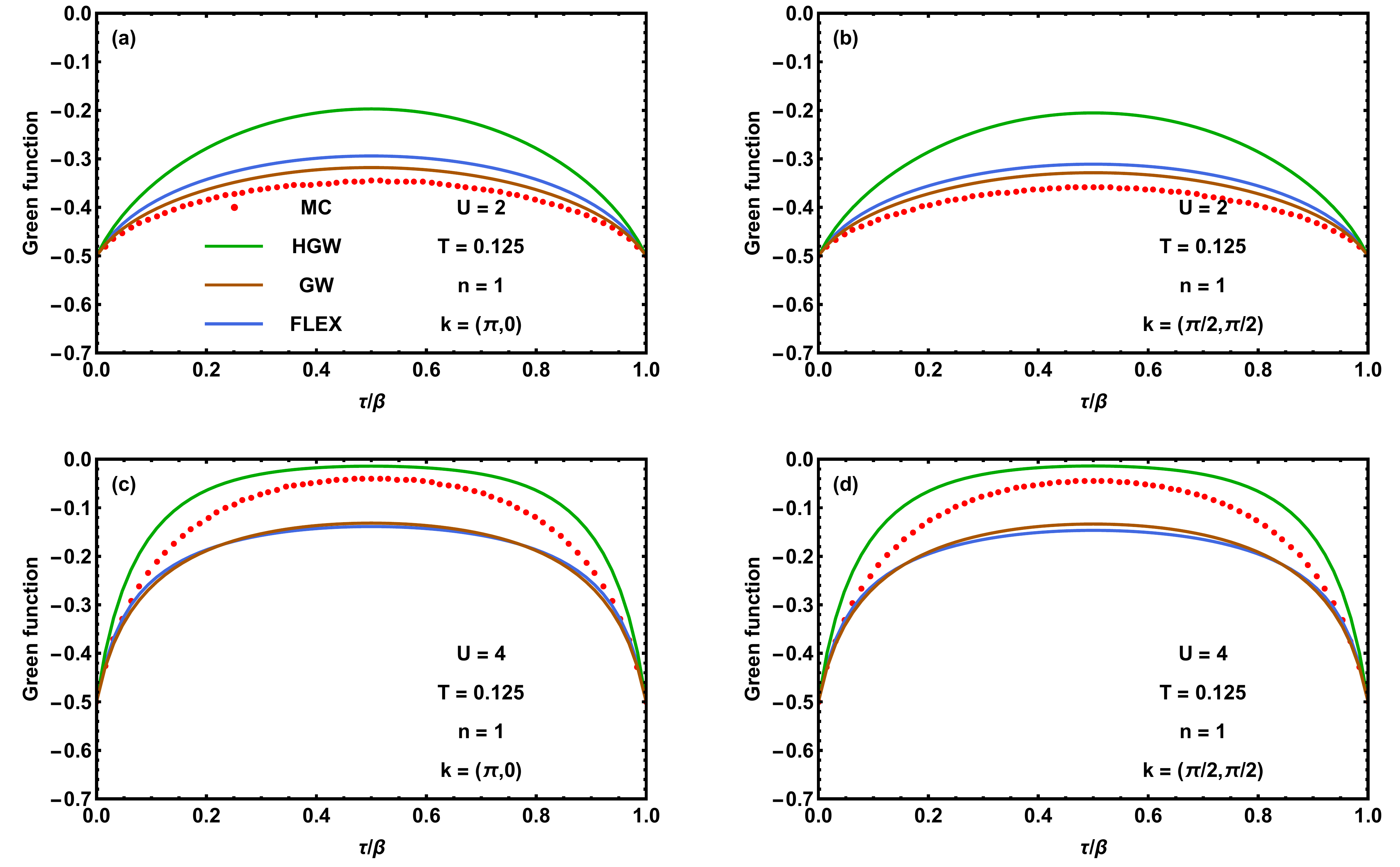}   
\caption{Comparison of results of Green's function at Matsubara time axis
for $8\times8$ cluster for different parameters: (a) $U=2,T=0.125,n=1,k=(%
\protect\pi,0)$, (b) $U=2,T=0.125,n=1,k=(\protect\pi/2,\protect\pi/2)$, (c) $%
U=4,T=0.125,n=1,k=(\protect\pi,0)$, (d) $U=4,T=0.125,n=1,k=(\protect\pi/2,%
\protect\pi/2)$ . The red dots denote the results obtained from
determinantal MC. The darker green line denotes the results obtained from $HGW$
equations. The darker orange line denotes the results obtained from $GW$
equations. The royal blue line denotes the results obtained from FLEX
approximation.}
\label{fig: comparison_green_func_8x8}
\end{figure}

These results can be understood as follows. The $HGW$ approximation is
obtained by truncating the three - body connected correlators, which have a
good clustering property at a stronger coupling $U$. At $U=2$, the three -
body connected correlators might be very nonlocal and the inequality (\ref%
{eq:inequality-hgw}) does not hold, and as a result, the $HGW$ method
preforms not so good. As a contrast, at $U=4$, the connected correlators
become local and the $HGW$ method exhibits its advantage. 


\subsubsection{Spectral function at half - filling}

\begin{figure}[tbph]
\includegraphics[width=\linewidth]{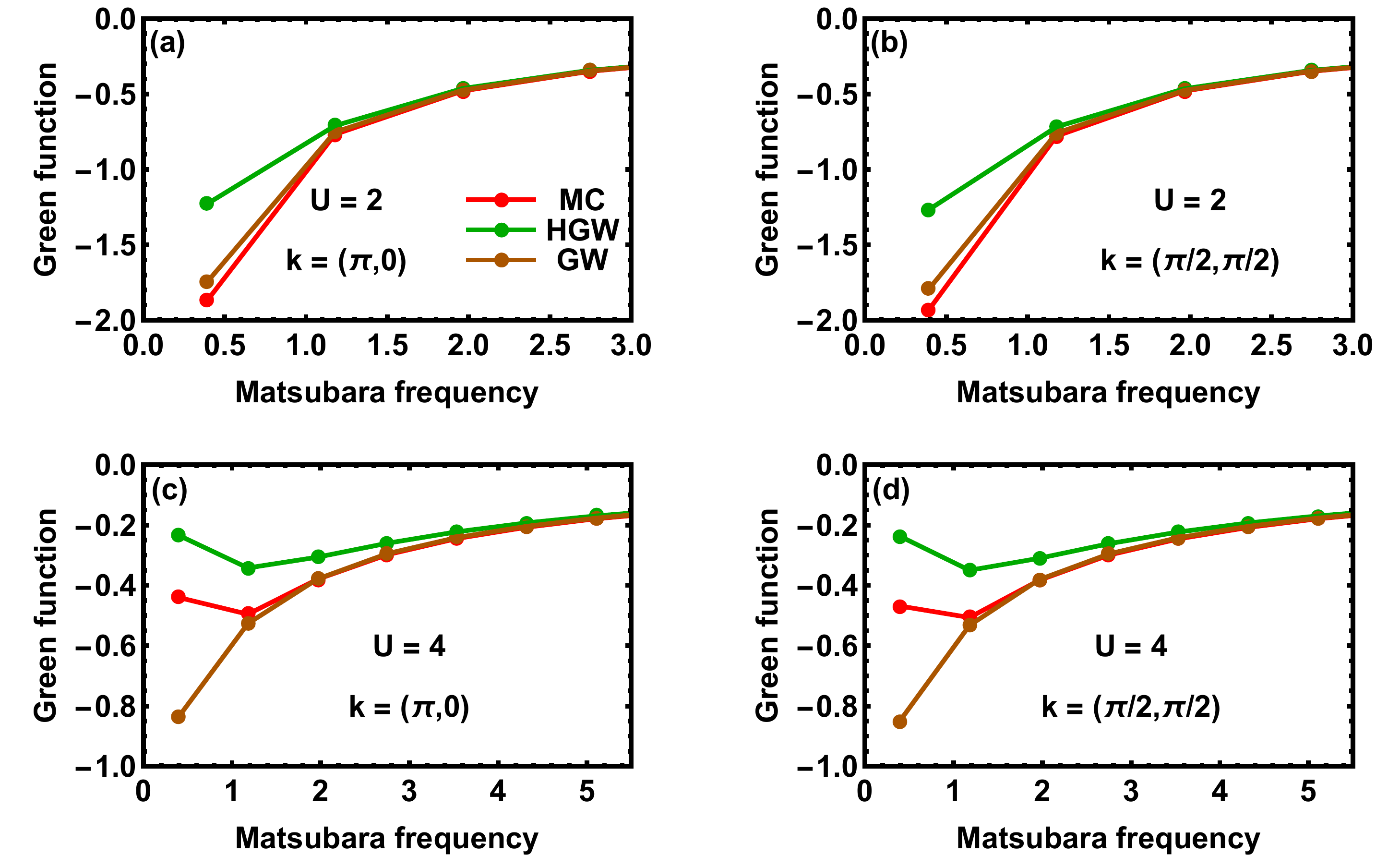}  
\caption{Comparison of the results of the imaginary part of the Green's
function at the Matsubara frequency axis for $8\times8$ cluster for
different parameters: (a) $U=2,T=0.125,n=1,k=(\protect\pi,0)$, (b) $%
U=2,T=0.125,n=1,k=(\protect\pi/2,\protect\pi/2)$, (c) $U=4,T=0.125,n=1,k=(%
\protect\pi,0)$, (d) $U=4,T=0.125,n=1,k=(\protect\pi/2,\protect\pi/2)$ . The
red dots denote the results obtained from determinantal MC. The darker green
line denotes the results obtained from $HGW$ equations. The darker orange line
denotes the results obtained from $GW$ equations.}
\label{fig: comparison_gfq_8x8}
\end{figure}

Using the discrete Fourier transformation, one obtains the values of the
Green's function at small Matsubara frequencies from those at Matsubara time
axis. The comparison of imaginary part of the value of the Green's function
at Matsubara frequency axis at half - filling for $8\times8$ cluster is
shown in Fig.\ref{fig: comparison_gfq_8x8}. These results demonstrate again
that the $HGW$ method is worse than the tranditional $GW$ method in the weak
coupling regime. At a stronger couling, $U=4$, the shape of the HGW curve
implies a Mott - Hubbard gap, just like the MC curve. On the contrary, the $GW$ method
fails.

\begin{figure}[tbph]
\includegraphics[width=\linewidth]{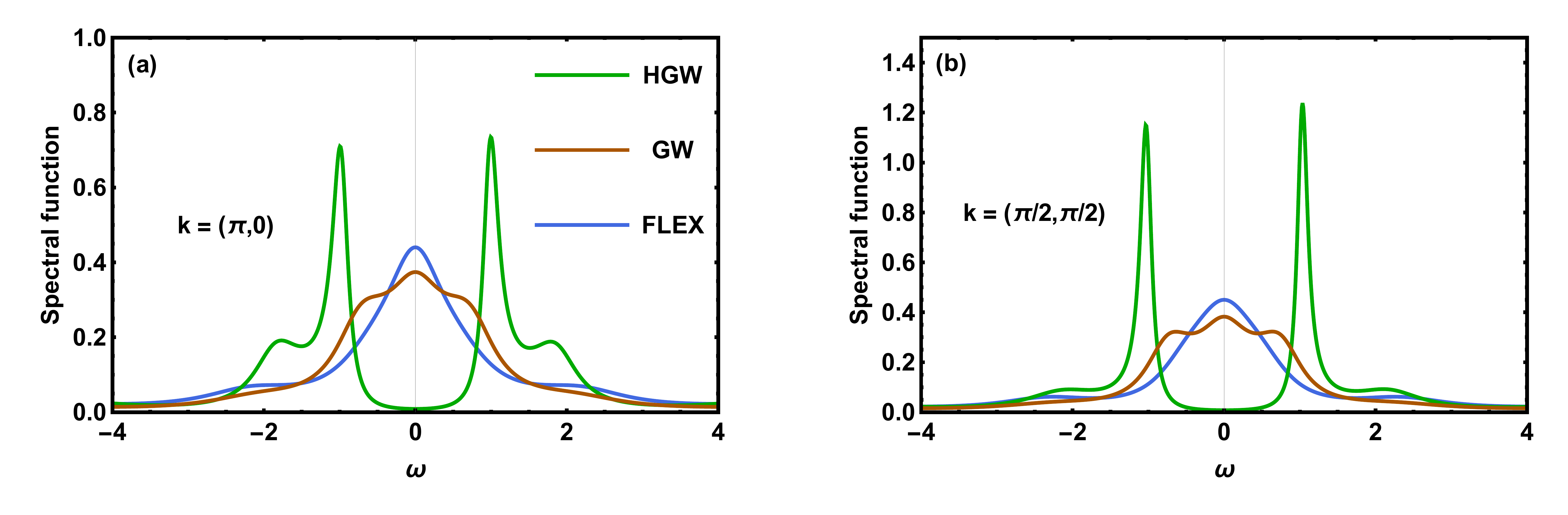}  
\caption{Comparison of the results of the spectral functions for $8\times8$
cluster at $U=4, T=0.125$ and half - filling at different momenta: (a) $k=(\protect\pi,0)$, 
(b) $k=(\protect\pi/2,\protect\pi/2)$. 
The darker green line denotes for the results of the
spectral function obtained from the $HGW$ method, the darker orange line
denotes for those obtained from the $GW$ method, and the royal blue line
denotes for those obtained from the FLEX theory.}
\label{fig: comparison_spfunc_8x8}
\end{figure}

With the values of the Green's function at some Matsubara frequencies, one
can obtain the spectral function by the analytical continuation. We adopt
the Nevanlinna analytical continuation[\onlinecite{fei_nevanlinna_2021}],
which is applicable to noiseless Matsubara data. The results of the spectral
function at $U=4, T=0.125$ at half - filling for the $8\times 8$ cluster 
are shown in Fig.\ref{fig: comparison_spfunc_8x8}. The spectral function obtained from
the $HGW$ method does exhibit a Mott - Hubbard gap. The spectral function for 2D half - filling 
Hubbard model has been studied by various methods, for example, the Monte Carlo simulation[\onlinecite{bulut1994}],
the ladder dual fermion approximation[\onlinecite{tanaka2019}], 
the celullar dynamical mean field theory[\onlinecite{kyung2006}], 
and the cluster perturbation theory[\onlinecite{senechal2000}]. 
We found that our results are similar to those obtained by the cluster perturbation theory 
(Fig.9c presented in Ref[\onlinecite{tanaka2019}]). 

\subsection{Charge density correlator and charge susceptibility at half
filling}

We compare the charge correlator in Matsubara time at the quasi - momentum $%
\left( \pi ,\pi \right) $ obtained from the covariant $HGW$ approximation ($%
\text{c}HGW$) with those based on the RPA formula (\ref{eq:chi_rpa_chi_0_rpa}%
, \ref{eq:solution_chi_0_rpa}), where the Green's functions $G$ obtained
from the $HGW$, $GW$, FLEX approximations are used.

\begin{figure}[tbph]
\includegraphics[width=\linewidth]{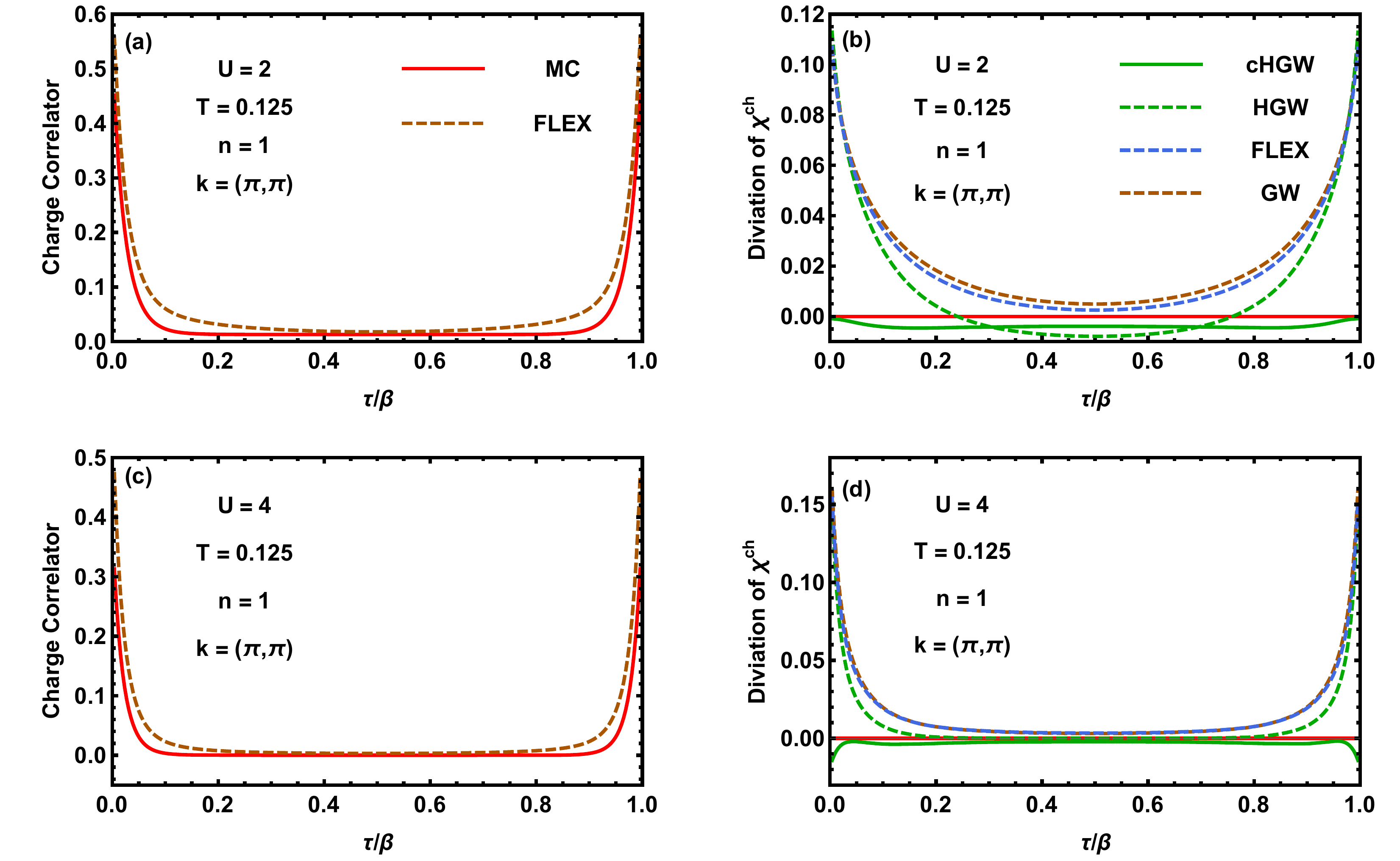}  
\caption{Comparison of results of charge correlator in Matsubara time at $%
k=( \protect\pi ,\protect\pi )$ the $4\times 4$ Hubbard cluster. For $%
U=2,T=0.125,n=1$, (a) shows the results of charge correlator obtained from
FLEX and MC, and (b) shows the differences between results of charge
correlator obtained from different approximations and those obtained from
MC. For $U=4,T=0.125,n=1$, (c) shows the results of charge correlator, and
(d) shows the differences. The red line denotes for MC. The darker green
solid line denotes for $\text{c}HGW$, and the darker green dashed line
denotes for $HGW$. The darker orange dashed line denotes for $GW$. The royal
blue dashed line denotes for FLEX. }
\label{fig: comparison_charge_correlator}
\end{figure}

We study the $4\times4$ cluster and set $M=1024$. Two sets of parameters are
chosen: $U=2,T=0.125$ in Figs.\ref{fig: comparison_charge_correlator}(a, b),
and $U=4,T=0.125$ in Figs.\ref{fig: comparison_charge_correlator}(c, d).
Since the results turn out to be too close to differentiate, only the FLEX
and MC curves for the charge correlator are plotted in Figs.\ref{fig:
comparison_charge_correlator}(a, c), and the differences between results
obtained from the above approximations and those obtained from MC are
plotted in Figs.\ref{fig: comparison_charge_correlator}(b, d).

In Fig.\ref{fig: comparison_charge_correlator}(b) with parameter $U=2,T=0.125
$, the largest differences given by $HGW$, $GW$ and FLEX (within the RPA
formula) are all about $0.12$ (near $\tau =0$ and $\tau =\beta $), while
that given by $\text{c}HGW$ is about $0.01$. In Fig.\ref{fig:
comparison_charge_correlator}(d) with parameter $U=4,T=0.125$, the largest
differences (near $\tau =0$ and $\tau =\beta $) given by $HGW$, $GW$ and
FLEX are all about $0.15$, while that given by $\text{c}HGW$ is less than $%
0.01$. These results demonstrate that the covariant scheme makes a
significant improvement over the RPA calculations.

We study the dependence of the static charge susceptibility $\chi^{\text{ch}%
}\left(i\Omega=0, k\right)$ on the coupling strength $U$ at $T=0.125$, and
these results are presented in Figs.\ref{fig:
comparison_static_charge_susceptibility}(a, b). The curves obtained from the
c$HGW$ method and the RPA calculations with the Green's function obtained
from the $HGW$, $GW$ and FLEX approximations have a similar tendency to the
MC curve. The c$HGW$ curve is much closer to the MC curve, which
demonstrates again the covariant scheme makes a significant improvement over
the RPA calculations.

\begin{figure}[tbph]
\includegraphics[width=\linewidth]{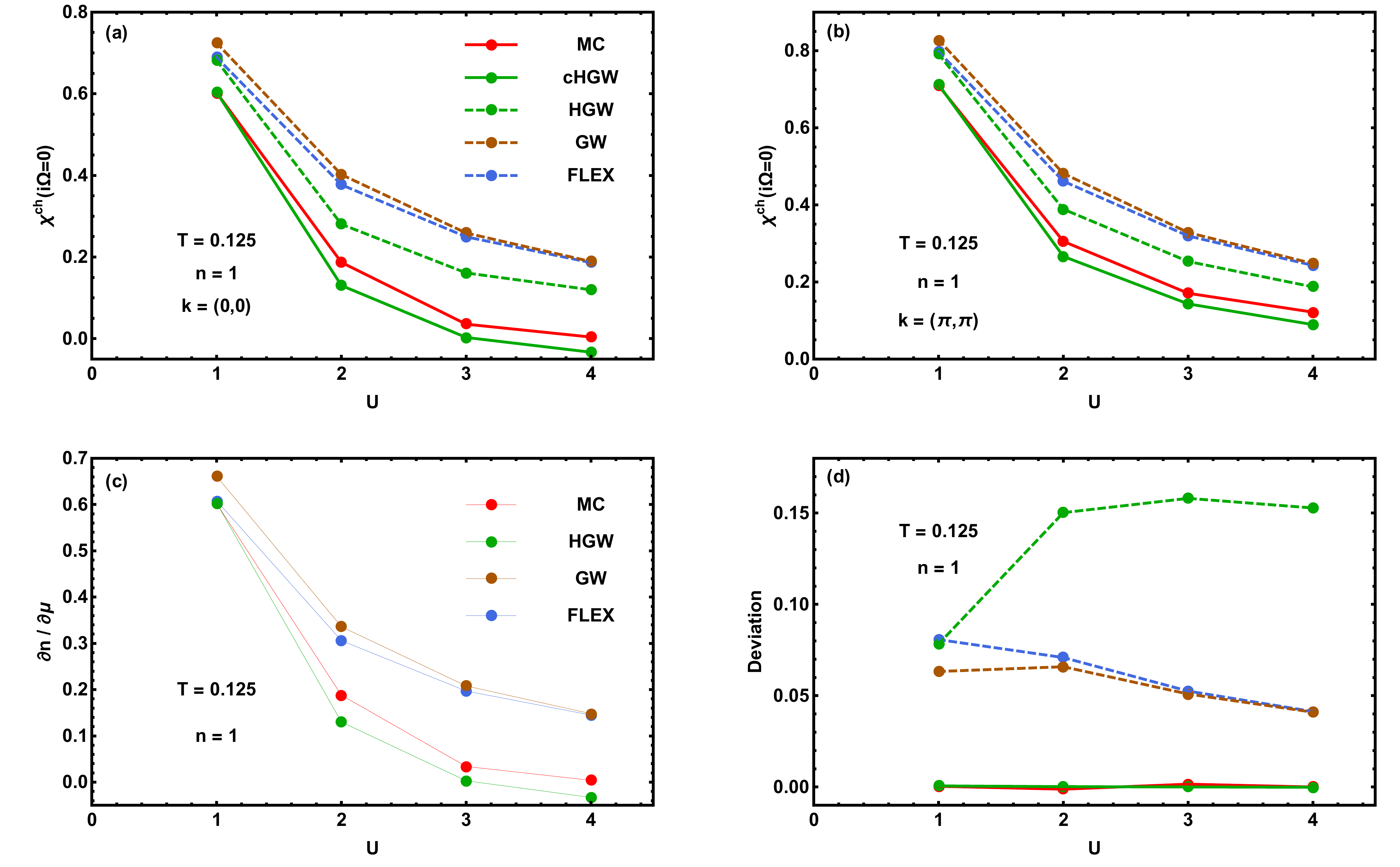}  
\caption{Comparison of the results of the static charge susceptibility
dependence of $U$ at $T=0.125$ for the $4\times 4$ Hubbard cluster. (a)
shows the results of the static charge susceptibility obtained from the MC, c%
$HGW$, $HGW$, $GW$, FLEX methods, at $k=(0,0)$, (b) at $k=(\protect\pi,%
\protect\pi)$. (c) shows the values of $\partial n / \partial \protect\mu$
obtained from the independent calculations through MC, $HGW$, $GW$ and FLEX
methods. (d) shows the deviation of the RPA calculations. In Figs.(a, b, d),
the red line denotes for MC, the darker green solid line denotes for $\text{c%
}HGW$, the darker green dashed line denotes for $HGW$, the darker orange
dashed line denotes for $GW$, and the royal blue dashed line denotes for
FLEX. In Fig.(c), the red line denotes for MC, the darker green solid line
denotes for $HGW$, the darker orange solid line denotes for $GW$, and the
royal blue solid line denotes for FLEX.}
\label{fig: comparison_static_charge_susceptibility}
\end{figure}

We also compare the values of $\partial n/\partial \mu $ (by variation of the denstiy with 
the chemical potential, i.e. $\Delta n / \Delta \mu$) at different couplings, 
and the results are presented in
Fig.\ref{fig: comparison_static_charge_susceptibility}(c). The $HGW$ curve
is much closer to the MC curve than the $GW$ and FLEX curves. The tendency
of $\partial n/\partial \mu $ to $0$ as $U$ increases showed by the MC results
demonstrates the Mott - Hubbard gap at strong coupling.

In a self - consistent theory, the static charge susceptibility $\chi_{\text{%
c}}\equiv\chi^{\text{ch}}\left(i\Omega=0, k=0\right)$ is equal to the
quantity $\partial n / \partial \mu$ from an independent calculation. To
study this consistency, we compare the quantity $\partial n / \partial \mu -
\chi_{\text{c}}$, and the results are presented in Fig.\ref{fig:
comparison_static_charge_susceptibility}(d). The results demonstrate that
the MC and the covariant calculations hold the consistency, whereas the RPA
calculations have significant deviations.

\section{Conclusion and Discussion \label{sec:conclusion}}

To summarize, the $HGW$ approximation, a modified $GW$ approximation, is
developed. It is derived by introduction of an external source $\phi $
coupled to the density $\rho $ and truncation of high order correlators on
equations of motion. The complexity of the $HGW$ equations turn out to be
very similar to $GW$ equations. The $HGW$ approximation is compared with
other approximations of comparable complexity $GW$, FLEX in the Hubbard
model. The results of the density and Green's function demonstrate that the $%
HGW$ approximation has a significant advantage over $GW$ and FLEX in a
relative strong coupling regime especially away from half - filling. 
More importantly, the $HGW$ approximation exhibits a gap as $U$ increases,
whereas the $GW$ and FLEX methods fail. 

To obtain the charge - conserving two - body correlators in the $HGW$
approximation, the covariant scheme is developed. In this scheme, the two -
body correlators are calculated through functional derivatives of Green's
function $G$ with respect to the source $\phi $. The covariant scheme for
the charge correlator is compared with the RPA scheme and determinantal MC
in the Hubbard model. The comparison demonstrates that the covariant scheme
makes a significant improvement over the RPA scheme. The comparison of
charge susceptibility demonstrates that the covariant scheme for charge
correlator is consistent with the charge susceptibility, whereas the RPA
calculation has a significant deviation.

The $HGW$ method (to calculate the one - body Green's function) has a small
complexity (for details, see Appendix \ref{sec:algorithm}), and thus can be
applied to large systems. The formalism presented in this paper is easily
extended to more general cases, such as multi - orbital lattice models, as
long as the band index is put in the generalized coordinate. The similarity
to the traditional $GW$ method helps the application of the $HGW$ method to
studying the electronic properties of realistic correlated materials. The
substantial improvement over the $GW$ method in relatively strong coupling
regime might imply that the $HGW$ method is a good alternative in certain
cases. However, the numerical cost of the calculation of the charge -
conserving charge correlators is too large for realistic systems.

To fully study the Hubbard model, the spin channel is important, whereas
neglected in our current formalism (the tranditional $GW$ method also
neglects the spin channel). The variant of $GW$ method including spin
channel were proposed for example in Ref.[\onlinecite{aryal_trilex_2015,
aryal_trilex_2016, vucicevic_trilex_2017}]. The variant of $HGW$
approximation could also be proposed by including spin channel to better
account the spin fluctuation at strong fluctuation regime in the 2D Hubbard
model in future research.

The self consistency is important to non - perturbative analytical methods,
and numerous ideas are put forward to ensure several identities. For
example, in the two - particle self - consistent (TPSC) theory [%
\onlinecite{vilk1997non, miyahara2013, zantout2021}], several ``constants'' are
determined by the sum rules and identities. In contrast, the idea of the
covariance is natural and universal in a sense that the correlators and the
sum rules are treated in the same footing.

\begin{acknowledgments}

    This work is supported by High-performance Computing Platform of Peking University. B.R. was supported
    by MOST of Taiwan, Grants No. 107-2112-M-003-023-MY3. D.P.L. was supported by National
    Natural Science Foundation of China, Grants No. 11674007 and No. 91736208.
    B.R. and D.P.L. are grateful to School of Physics of Peking University and The Center for Theoretical Sciences of Taiwan for hospitality, respectively.

\end{acknowledgments}

\appendix

\section{Dyson - Schwinger equations and Ward identities}

\subsection{Dyson-Schwinger equations of motion \label{sec:derivation-eom}}

The invariance of the functional integral measure $\mathcal{D}\left[ \psi
,\psi ^{\ast }\right] $ under the infinitesimal variation of field $\psi
,\psi ^{\ast } $ yields the equality [\onlinecite{peskin_introduction_2018}] 
\begin{equation}
\int \mathcal{D}\left[ \psi ,\psi ^{\ast }\right] \ \frac{\delta }{\delta
\psi ^{\ast }\left( 2\right) }\left( \psi ^{\ast }\left( 1\right) \mathrm{e}%
^{-S\left[ \psi ,\psi ^{\ast };\phi \right] }\right) =0.
\label{eq:ds-equality}
\end{equation}%
Substituting the perturbed action \eqref{eq:perturbed_action} into the
equality, one obtains the Dyson - Schwinger equation of motion: 
\begin{equation}
\delta \left( 1,2\right) =\int d\left( 3\right) \ T\left( 1,3\right) G\left(
3,2\right) +\phi \left( 1\right) G\left( 1,2\right) -\int d\left( 3\right) \
V\left( 1,3\right) G_{2}\left( 1,2;3,3\right) .  \label{eq:ds-equation}
\end{equation}%
Here the two - body correlator is defined by 
\begin{align}
G_{2}\left( 1,2;3,4\right) & \equiv \left \langle \psi ^{\ast }\left(
2\right) \psi \left( 1\right) \psi ^{\ast }\left( 4\right) \psi \left(
3\right) \right \rangle  \notag \\
& =\frac{1}{Z\left[ \phi \right] }\int \mathcal{D}\left[ \psi ,\psi ^{\ast }%
\right] \ \psi ^{\ast }\left( 2\right) \psi \left( 1\right) \psi ^{\ast
}\left( 4\right) \psi \left( 3\right) \text{e}^{-S\left[ \psi ,\psi ^{\ast
};\phi \right] }.  \label{eq:definition_of_two_body_correlator}
\end{align}

Through the definition \eqref{eq:definition-of-green-function}, one obtains
the derivative of $G$ with respect to $\phi $: 
\begin{equation}
L\left( 1,2;3\right) \equiv \frac{\delta G\left( 1,2\right) }{\delta \phi
\left( 3\right) }=G_{2}\left( 1,2;3,3\right) -G\left( 1,2\right) \rho \left(
3\right) ,  \label{eq:dg-in-g2}
\end{equation}%
where $\rho \left( 1\right) \equiv \left \langle \rho \left( 1\right)
\right
\rangle =G\left( 1,1\right) $. By virtue of Eq.\eqref{eq:dg-in-g2},
one can express $G_{2}$ in terms of $G$ and $\delta G/\delta \phi $, and
thus can obtain Eq.\eqref{eq:equation-of-motion} from Eq.%
\eqref{eq:ds-equation}.

\subsection{Ward identities for correlators \label{sec:ward-identity}}

The invariance of the functional integral measure $\mathcal{D}\left[ \psi
,\psi ^{\ast }\right] $ under the infinitesimal phase rotation of the
complex field $\psi $ yields an equality [%
\onlinecite{peskin_introduction_2018}] 
\begin{equation}
\int \mathcal{D}\left[ \psi ,\psi ^{\ast }\right] \ \left( \psi ^{\ast
}\left( 1\right) \frac{\delta }{\delta \psi ^{\ast }\left( 1\right) }-\psi
\left( 1\right) \frac{\delta }{\delta \psi \left( 1\right) }\right) \mathrm{e%
}^{-S\left[ \psi ,\psi ^{\ast };\phi \right] }=0.  \label{eq:ward-equality}
\end{equation}%
Substituting the perturbed action (\ref{eq:perturbed_action}), one obtains
the Ward identity for Green's function $G$: 
\begin{equation}
\int d\left( 2\right) \ T\left( 1,2\right) G\left( 2,1\right) -T\left(
2,1\right) G\left( 1,2\right) =0.  \label{eq:ward-identity-g}
\end{equation}%
The derivative of Eq.\eqref{eq:ward-identity-g} with respect to $\phi $
yields: 
\begin{equation}
\int d\left( 2\right) \ T\left( 1,2\right) L\left( 2,1;3\right) -T\left(
2,1\right) L\left( 1,2;3\right) =0.  \label{eq:ward-identity-l}
\end{equation}%
Eq.\eqref{eq:ward-identity-l} is the Ward identity for the two - body
correlator $L$.

\section{Details of deriving $HGW$ equations and covariant $HGW$ equations}

\label{sec:derivation-hgw}

\subsection{Derivation of $HGW$ equations}

\label{sec:derivation-hgw-equations} The $HGW$ equations are derived from
Eqs.(\ref{eq:equation-of-motion}, \ref{eq:hgw-approximation}). First, one
makes derivative of Eqs.(\ref{eq:definition_of_h}, \ref{eq:definition_of_v})
with respect to $\phi $, and obtains 
\begin{equation}
\frac{\delta H^{-1}\left( 1,2\right) }{\delta \phi \left( 3\right) }=\delta
\left( 1,2\right) \frac{\delta v\left( 1\right) }{\delta \phi \left(
3\right) },  \label{eq:deltah_deltaphi}
\end{equation}%
and 
\begin{equation}
\frac{\delta v\left( 1\right) }{\delta \phi \left( 2\right) }=\delta \left(
1,2\right) -\int d\left( 3\right) \ V\left( 1,3\right) \frac{\delta \rho
\left( 3\right) }{\delta \phi \left( 2\right) }.  \label{eq:deltav_deltaphi}
\end{equation}%
Substituting Eq.(\ref{eq:deltah_deltaphi}) into Eq.(\ref%
{eq:hgw-approximation}) leads to 
\begin{equation}
\frac{\delta G\left( 1,2\right) }{\delta \phi \left( 3\right) }=-\int
d\left( 4\right) \ H\left( 1,4\right) G\left( 4,2\right) \frac{\delta
v\left( 4\right) }{\delta \phi \left( 3\right) }.
\label{eq:deltag_deltaphi_in_deltav_deltaphi}
\end{equation}%
Plugging Eq.(\ref{eq:deltag_deltaphi_in_deltav_deltaphi}) into Eq.(\ref%
{eq:deltav_deltaphi}), and one obtains 
\begin{equation}
\frac{\delta v\left( 1\right) }{\delta \phi \left( 2\right) }=\delta \left(
1,2\right) +\int d\left( 34\right) \ V\left( 1,3\right) \Pi \left(
3,4\right) \frac{\delta v\left( 4\right) }{\delta \phi \left( 2\right) },
\label{eq:delta_v_deltaphi_in_pi}
\end{equation}%
with 
\begin{equation}
\Pi \left( 1,2\right) \equiv H\left( 1,2\right) G\left( 2,1\right) .
\label{eq:definition_of_polarization}
\end{equation}

Then substituting Eq.(\ref{eq:deltag_deltaphi_in_deltav_deltaphi}) into Eq.%
\eqref{eq:equation-of-motion}, one obtains 
\begin{equation}
\delta \left( 1,2\right) =\int d\left( 3\right) \ H^{-1}\left( 1,3\right)
G\left( 3,2\right) +\int d\left( 34\right) \ V\left( 1,3\right) H\left(
1,4\right) G\left( 4,2\right) \frac{\delta v\left( 4\right) }{\delta \phi
\left( 3\right) }.  \label{eq:1st_eom_in_deltav_deltaphi}
\end{equation}%
The equation above can be rewritten as 
\begin{equation}
G^{-1}\left( 1,2\right) =H^{-1}\left( 1,2\right) -\Sigma \left( 1,2\right) ,
\label{eq:dyson_equation}
\end{equation}%
with the self - energy function $\Sigma$ given by 
\begin{equation}
\Sigma \left( 1,2\right) \equiv -H\left( 1,2\right) W\left( 2,1\right) ,
\label{eq:definition_of_sigma}
\end{equation}%
and the screened dynamical potential $W$ defined by 
\begin{equation}
W\left( 1,2\right) \equiv \int d\left( 3\right) \ \frac{\delta v\left(
1\right) }{\delta \phi \left( 3\right) }V\left( 2,3\right) .
\label{eq:definition_of_w}
\end{equation}

Combining Eqs.(\ref{eq:delta_v_deltaphi_in_pi}, \ref{eq:definition_of_w}),
one arrives at the following equation 
\begin{equation}
W\left( 1,2\right) =V\left( 1,2\right) +\int d\left( 34\right) \ V\left(
1,3\right) \Pi \left( 3,4\right) W\left( 4,2\right) ,
\label{eq:equation_of_w}
\end{equation}%
which can be rewritten as 
\begin{equation}
W^{-1}\left( 1,2\right) =V^{-1}\left( 1,2\right) -\Pi \left( 1,2\right) .
\label{eq:equation_of_w_inv}
\end{equation}
Now four important equations (\ref{eq:definition_of_polarization}, \ref%
{eq:dyson_equation}, \ref{eq:definition_of_sigma}, \ref{eq:equation_of_w_inv}%
) are derived and they are called the $HGW$ equations.

\subsection{Derivation of covariant $HGW$ equations \label%
{sec:derivation-covariant-hgw}}

Here, details of derivation of covariant $HGW$ equations are given. For
convenience, one can introduce the covariant versions of two vertex
functions: 
\begin{equation}
\Lambda \left( 1,2;3\right) \equiv \left.\frac{\delta G^{-1}\left(
1,2\right) }{\delta v\left( 3\right) }\right \vert_{\phi=0},
\label{eq:definition_of_lambda}
\end{equation}%
and 
\begin{equation}
\Gamma \left( 1,2;3\right) \equiv \left.\frac{\delta W^{-1}\left( 1,2\right) 
}{\delta v\left( 3\right) }\right \vert_{\phi=0}.
\label{eq:definition_of_gamma}
\end{equation}
The derivatives of $HGW$ equations with respect to $v$ can be easily
obtained 
\begin{equation}
\Lambda \left( 1,2;3\right) =\frac{\delta H^{-1}\left( 1,2\right) }{\delta
v\left( 3\right) }+\frac{\delta H\left( 1,2\right) }{\delta v\left( 3\right) 
}W\left( 2,1\right) +H\left( 1,2\right) \frac{\delta W\left( 2,1\right) }{%
\delta v\left( 3\right) },  \label{eq:derivative-equation-lambda-1}
\end{equation}%
\begin{equation}
\Gamma \left( 1,2;3\right) =-\frac{\delta H\left( 1,2\right) }{\delta
v\left( 3\right) }G\left( 2,1\right) -H\left( 1,2\right) \frac{\delta
G\left( 2,1\right) }{\delta v\left( 3\right) },
\label{eq:derivative-equation-gamma-1}
\end{equation}%
There is a general relation for an array $X$: 
\begin{equation}
\frac{\delta X\left( 1,2\right) }{\delta v\left( 3\right) }=-\int d\left(
45\right) \ X\left( 1,4\right) X\left( 5,2\right) \frac{\delta X^{-1}\left(
4,5\right) }{\delta v\left( 3\right) },  \label{eq:deltax_deltaxinv}
\end{equation}%
and for $X=H$, 
\begin{equation}
\frac{\delta H^{-1}\left( 1,2\right) }{\delta v\left( 3\right) }=\delta
\left( 1,2\right) \delta \left( 1,3\right).  \label{eq:delta_hinv_delta_v}
\end{equation}

Then one can obtain the equation for $\Lambda $ 
\begin{align}
\Lambda \left( 1,2;3\right) & =\delta \left( 1,2\right) \delta \left(
1,3\right) -H\left( 1,3\right) H\left( 3,2\right) W\left( 2,1\right)  \notag
\\
& \quad -\int d\left( 45\right) \ H\left( 1,2\right) W\left( 2,5\right)
W\left( 4,1\right) \Gamma \left( 5,4;3\right) ,
\label{eq:derivative-equation-lambda}
\end{align}%
and the equation for $\Gamma $ 
\begin{align}
\Gamma \left( 1,2;3\right) & =H\left( 1,3\right) H\left( 3,2\right) G\left(
2,1\right)  \notag \\
& \quad +\int d\left( 45\right) \ H\left( 1,2\right) G\left( 2,5\right)
G\left( 4,1\right) \Lambda \left( 5,4;3\right) .
\label{eq:derivative-equation-gamma}
\end{align}%
From Eqs (\ref{eq:derivative-equation-lambda}, \ref%
{eq:derivative-equation-gamma}), one can obtain $\Lambda $ and $\Gamma $,
giving $H,G,W$ obtained from on - shell ($\phi =0$) $HGW$ equations.

One can introduce $\chi_{0}$ as 
\begin{equation}
\chi _{0}\left( 1,2\right) \equiv \left. \frac{\delta \rho \left( 1\right) }{%
\delta v\left( 2\right) }\right \vert_{\phi=0},  \label{eq:chi-zero-def}
\end{equation}%
and then obtains Eq.(\ref{eq:chi-zero-cov-equation}). With the definitions (%
\ref{eq:chi_cov_definition}, \ref{eq:chi-zero-def}), one obtains 
\begin{equation}
\chi ^{\text{cov}}\left( 1,2\right) =\int d\left( 3\right) \ \chi _{0}\left(
1,3\right) \frac{\delta v\left( 3\right) }{\delta \phi \left( 2\right) }.
\label{eq:chi_cov_chi_0_cov_deltav_deltaphi}
\end{equation}%
By virtue of Eq.(\ref{eq:deltav_deltaphi}), Eq.(\ref%
{eq:chi_cov_chi_0_cov_deltav_deltaphi}) leads to Eq.(\ref%
{eq:chi-cov-chi-zero}).

\section{$HGW$ and covariant $HGW$ equations for the Hubbard model}

\subsection{Generalized Fourier transformation for the Hubbard model}

\label{sec:fourier-hubbard} The generalized Fourier transformation for the
Hubbard model is introduced here. For a short formulation, two useful
notations 
\begin{equation}
\mathcal{E}_{\text{F}}\left( \alpha ,1-2\right) \equiv \text{e}^{\text{i}\pi
\eta _{\alpha }\cdot \left( \sigma _{1}-\sigma _{2}\right) }\text{e}^{\text{i%
}k_{\alpha }\cdot \left( i_{1}-i_{2}\right) }\text{e}^{\text{i}\pi \left(
2m_{\alpha }+1\right) \cdot \left( \tau _{1}-\tau _{2}\right) },
\label{eq:fermionic_fourier_transformation}
\end{equation}%
\begin{equation}
\mathcal{E}_{\text{B}}\left( \alpha ,1-2\right) \equiv \text{e}^{\text{i}\pi
\eta _{\alpha }\cdot \left( \sigma _{1}-\sigma _{2}\right) }\text{e}^{\text{i%
}k_{\alpha }\cdot \left( i_{1}-i_{2}\right) }\text{e}^{\text{i}\pi
2m_{\alpha }\cdot \left( \tau _{1}-\tau _{2}\right) },
\label{eq:bosonic_fourier_transformation}
\end{equation}%
are introduced, where the label $\alpha $ refers to $\eta _{\alpha
},k_{\alpha },m_{\alpha }$. For a fermionic array $X_{\text{F}}$ which is
anti - periodic over Matsubara time, one can expand it as Fourier series: 
\begin{equation}
X_{\text{F}}\left( 1,2\right) =\frac{1}{\mathcal{N}}\sum_{\alpha }\tilde{X}_{%
\text{F}}\left( \alpha \right) \mathcal{E}_{\text{F}}\left( \alpha
,1-2\right) .  \label{eq:fermionic-ansatz}
\end{equation}%
Here $\mathcal{N}=2MN^{2}$, $M$ is the number of time slices and $N^{2}$ is
the number of lattice sites. $\sigma $ is quantified as $1$ for spin-down
and $0$ for spin-up, and correspondingly $\eta $ takes value of $0$ or $1$. $%
i$ is the coordinate of lattice site, and $k$ is the momentum in the first
Brillouin zone. $\tau \in \left[ 0,\beta \right] $ is the discrete Matsubara
time, and $m$ takes integral value from $0$ to $M-1$. The summation $%
\sum_{\alpha }$ is over all possible values of $\alpha \equiv \left( \eta
_{\alpha },k_{\alpha },m_{\alpha }\right) $. Similarly, one can expand a
bosonic array $X_{\text{B}}$, which is periodic over Matsubara time, as
Fourier series: 
\begin{equation}
X_{\text{B}}\left( 1,2\right) =\frac{1}{\mathcal{N}}\sum_{\alpha }\tilde{X}_{%
\text{B}}\left( \alpha \right) \mathcal{E}_{\text{B}}\left( \alpha
,1-2\right) .  \label{eq:bosonic-ansatz}
\end{equation}

The coefficient $T$ in action \eqref{eq:general-action} is anti - periodic
over Matsubara time and thus is a fermionic array. Substitute Eq.%
\eqref{eq:expression_for_t} into the ansatz \eqref{eq:fermionic-ansatz}, and
one obtains 
\begin{equation}
\tilde{T}\left( \alpha \right) =\Delta \tau \left( -\frac{1}{\Delta \tau }%
\left( \text{e}^{-\text{i}\pi \left( 2m+1\right) /M}-1\right)
-\varepsilon\left( k_{\alpha }\right) +\mu \right) ,
\label{eq:exp-t-fourier-space}
\end{equation}%
with 
\begin{equation}
\varepsilon \left( k\right) =-2t\left( \cos \left( k_{x}\right) +\cos \left(
k_{y}\right) \right) ,
\end{equation}%
for the 2D Hubbard model, with $k\equiv \left( k_{x},k_{y}\right) $. The
coefficient $V$ is periodic over Matsubara time and thus is a bosonic array.
Substituting Eq.\eqref{eq:expression_for_v} into the ansatz %
\eqref{eq:bosonic-ansatz}, one obtains 
\begin{equation}
\tilde{V}\left( \alpha \right) =\Delta \tau U\left( -1\right) ^{\eta
_{\alpha }}.  \label{eq:exp-v-fourier-space}
\end{equation}

From definitions (\ref{eq:fermionic_fourier_transformation}, \ref%
{eq:bosonic_fourier_transformation}), one can derive the following relations 
\begin{align}
\mathcal{E}_{\text{F}}\left( \alpha ,1-2\right) \mathcal{E}_{\text{F}}\left(
\alpha ,2-3\right) & =\mathcal{E}_{\text{F}}\left( \alpha ,1-3\right) , 
\notag \\
\mathcal{E}_{\text{B}}\left( \alpha ,1-2\right) \mathcal{E}_{\text{B}}\left(
\alpha ,2-3\right) & =\mathcal{E}_{\text{B}}\left( \alpha ,1-3\right) , 
\notag \\
\mathcal{E}_{\text{F}}\left( \alpha ,1-2\right) \mathcal{E}_{\text{B}}\left(
\beta ,1-2\right) & =\mathcal{E}_{\text{F}}\left( \alpha +\beta ,1-2\right) ,
\notag \\
\mathcal{E}_{\text{B}}\left( \alpha ,1-2\right) \mathcal{E}_{\text{B}}\left(
\beta ,1-2\right) & =\mathcal{E}_{\text{B}}\left( \alpha +\beta ,1-2\right) ,
\notag \\
\mathcal{E}_{\text{F}}\left( \alpha ,1-2\right) \mathcal{E}_{\text{F}}\left(
\beta ,2-1\right) & =\mathcal{E}_{\text{B}}\left( \alpha -\beta ,1-2\right) ,
\notag \\
\mathcal{E}_{\text{B}}\left( \alpha ,1-2\right) & =\mathcal{E}_{\text{B}%
}\left( -\alpha ,2-1\right) .  \label{eq:fourier_multiplication}
\end{align}%
These equations are helpful in the derivation of $HGW$ equations in Fourier
space.

\subsection{$HGW$ equations for the Hubbard model}

\label{sec:derivation-hgw-equation-hubbard} In $HGW$ equations, one
encounters with several quantities: the fermionic arrays $H, G, \Sigma$ and
the bosonic arrays $W, \Pi$. Substitute the ansatz (\ref{eq:fermionic-ansatz}%
, \ref{eq:bosonic-ansatz}) into the $HGW$ equations, and one obtains the $HGW
$ equations in Fourier space for the Hubbard model 
\begin{align}
\tilde{G}^{-1}\left(\alpha \right) & =\tilde{H}^{-1}\left(\alpha \right)-%
\tilde{\Sigma}\left(\alpha \right),  \notag \\
\tilde{\Sigma}\left(\alpha \right) & =-\frac{1}{\mathcal{N}}\sum_{\gamma}%
\tilde{H}\left(\alpha+\gamma \right)\tilde{W}\left(\gamma \right),  \notag \\
\tilde{W}^{-1}\left(\alpha \right) & =\tilde{V}^{-1}\left(\alpha \right)-%
\tilde{\Pi}\left(\alpha \right),  \notag \\
\tilde{\Pi}\left(\alpha \right) & =\frac{1}{\mathcal{N}}\sum_{\gamma}\tilde{H%
}\left(\alpha+\gamma \right)\tilde{G}\left(\gamma \right),
\label{eq:hgw_equations_in_fourier_space}
\end{align}
with 
\begin{equation}
\tilde{H}^{-1}\left(\alpha \right)=\tilde{T}\left(\alpha \right)-\frac{2U}{%
\mathcal{N}}\sum_{\gamma}\tilde{G}\left(\gamma \right).
\label{eq:h_equation_fourier_space}
\end{equation}

\subsection{Covariant $HGW$ equations for the Hubbard model\label%
{sec:derivation-covariant-hgw-hubbard}}

Similarly, one can obtain the covariant $HGW$ equations in Fourier space for
the Hubbard model. One can make the ansatz for the vertex functions, 
\begin{equation}
\Lambda \left(1,2,3\right)=\frac{1}{\mathcal{N}^2}\sum_{\alpha,\gamma}\tilde{%
\Lambda}\left(\alpha,\gamma \right)\mathcal{E}_{\text{F}}\left(\alpha,1-2%
\right)\mathcal{E}_{\text{B}}\left(\gamma,1-3\right),
\label{eq:ansatz_for_lambda}
\end{equation}
\begin{equation}
\Gamma \left(1,2,3\right)=\frac{1}{\mathcal{N}^2}\sum_{\alpha,\gamma}\tilde{%
\Gamma}\left(\alpha,\gamma \right)\mathcal{E}_{\text{B}}\left(\alpha,1-2%
\right)\mathcal{E}_{\text{B}}\left(\gamma,1-3\right).
\label{eq:ansatz_for_gamma}
\end{equation}
And one can obtain from Eqs.(\ref{eq:derivative-equation-lambda}, \ref%
{eq:derivative-equation-gamma}), 
\begin{align}
\tilde{\Lambda}\left(\alpha,\beta \right) &= 1-\frac{1}{\mathcal{N}}%
\sum_{\gamma}\tilde{H}\left(\alpha+\beta+\gamma \right)\tilde{H}%
\left(\alpha+\gamma \right)\tilde{W}\left(\gamma \right)  \notag \\
&\quad -\frac{1}{\mathcal{N}}\sum_{\gamma}\tilde{H}\left(\alpha+\beta+\gamma
\right)\tilde{W}\left(\beta+\gamma \right)\tilde{W}\left(\gamma \right)%
\tilde{\Gamma}\left(\gamma,\beta \right),  \label{eq:equation_of_lambda}
\end{align}
and 
\begin{align}
\tilde{\Gamma}\left(\alpha,\beta \right) &= \frac{1}{\mathcal{N}}%
\sum_{\gamma}\tilde{H}\left(\alpha+\beta+\gamma \right)\tilde{H}%
\left(\alpha+\gamma \right)\tilde{G}\left(\gamma \right)  \notag \\
&\quad +\frac{1}{\mathcal{N}}\sum_{\gamma}\tilde{H}\left(\alpha+\beta+\gamma
\right)\tilde{G}\left(\beta+\gamma \right)\tilde{G}\left(\gamma \right)%
\tilde{\Lambda}\left(\gamma,\beta \right).  \label{eq:equation_of_gamma}
\end{align}
Combine Eqs.(\ref{eq:equation_of_lambda}, \ref{eq:equation_of_gamma}), and
one obtains 
\begin{equation}
\sum_{\gamma}\mathcal{M}\left(\alpha,\gamma,\beta \right)\tilde{\Lambda}%
\left(\gamma,\beta \right)=b\left(\alpha,\beta \right),
\label{eq:lambda_equation_in_linear_form}
\end{equation}
with 
\begin{align}
\mathcal{M}\left(\alpha,\gamma,\beta \right) &= \delta(\alpha,\gamma) + 
\frac{1}{\mathcal{N}}\tilde{G}\left(\beta+\gamma \right)\tilde{G}%
\left(\gamma \right)  \notag \\
&\quad \times \frac{1}{\mathcal{N}}\sum_{\lambda}\tilde{H}%
\left(\alpha+\beta+\lambda \right)\tilde{W}\left(\beta+\lambda \right)\tilde{%
W}\left(\lambda \right)\tilde{H}\left(\beta+\gamma+\lambda \right),
\label{eq:matrix_in_linear_equation}
\end{align}
and 
\begin{align}
b\left(\alpha,\beta \right)=&1-\frac{1}{\mathcal{N}}\sum_{\gamma}\tilde{H}%
\left(\alpha+\beta+\gamma \right)\tilde{H}\left(\alpha+\gamma \right)\tilde{W%
}\left(\gamma \right)  \notag \\
&-\frac{1}{\mathcal{N}}\sum_{\gamma}\tilde{H}\left(\alpha+\beta+\gamma
\right)\tilde{W}\left(\beta+\gamma \right)\tilde{W}\left(\gamma \right) 
\notag \\
&\quad \times \frac{1}{\mathcal{N}} \sum_{\lambda}\tilde{H}%
\left(\beta+\gamma+\lambda \right)\tilde{H}\left(\gamma+\lambda \right)%
\tilde{G}\left(\lambda \right).  \label{eq:vector_in_linear_form}
\end{align}
Once Eq.(\ref{eq:lambda_equation_in_linear_form}) is solved, $\tilde{\Lambda}
$ will be obtained.

Next, one can make ansatz 
\begin{equation}
\chi_{0}\left(1,2\right)=\frac{1}{\mathcal{N}}\sum_{\alpha}\tilde{\chi}%
_{0}\left(\alpha \right)\mathcal{E}_{\text{B}}\left(\alpha,1-2\right),
\label{eq:chi-zero-def_fourier_ansatz}
\end{equation}
\begin{equation}
\chi^{\text{cov}}\left(1,2\right)=\frac{1}{\mathcal{N}}\sum_{\alpha}\tilde{%
\chi}^{\text{cov}}\left(\alpha \right)\mathcal{E}_{\text{B}%
}\left(\alpha,1-2\right).  \label{eq:chi_cov_fourier_ansatz}
\end{equation}
Then Eq.(\ref{eq:chi-zero-cov-equation}) yields 
\begin{equation}
\tilde{\chi}_{0}\left(\alpha \right)=-\sum_{\gamma}\tilde{G}%
\left(\alpha+\gamma \right)\tilde{G}\left(\gamma \right)\tilde{\Lambda}%
\left(\gamma,\alpha \right),  \label{eq:chi-zero-def_fourier_result}
\end{equation}
and Eq.(\ref{eq:chi-cov-chi-zero}) yields 
\begin{equation}
\tilde{\chi}^{\text{cov}}\left(\alpha \right)=\frac{\tilde{\chi}_{0}
\left(\alpha \right)}{1+\tilde{V}\left(\alpha \right)\tilde{\chi}_{0}
\left(\alpha \right)}.  \label{eq:chi_cov_fourier_result}
\end{equation}

Up to now, the covariant $HGW$ equations (\ref%
{eq:lambda_equation_in_linear_form}, \ref{eq:chi-zero-def_fourier_result}, %
\ref{eq:chi_cov_fourier_result}) are obtained in Fourier space for the
Hubbard model.

\section{Correlators in Matsubara action \label{sec:matsubara-action}}

The Matsubara action given in \eqref{eq:hubbard-action} is dependent of the
number of time slices $M$, and tends to the continuous time limit 
\begin{align}
S\left[ \psi ,\psi ^{\ast }\right] =\sum_{i,\sigma }\int_{0}^{\beta} d\tau \
\psi_{i,\sigma}^{\ast}\left(\tau \right) \partial_{\tau }
\psi_{i,\sigma}\left(\tau \right) +\int_{0}^{\beta} d\tau \ \mathcal{H} %
\left[ \psi_{i\sigma}^{\ast }\left(\tau \right) ,\psi_{i\sigma} \left(\tau
\right) \right] ,  \label{eq:action_continuous_limit}
\end{align}%
with a convergence speed $1/M$. For a short formulation, the spin and space
coordinates are dropped below. One can define the $M$-dependent Green's
function as 
\begin{equation}
G_{M}\left( \tau _{l_{1}}, \tau_{l_{2}}\right) \equiv \frac{1}{Z_{M} }\int 
\mathcal{D} \left[ \psi ,\psi ^{\ast}\right] \ \psi ^{\ast} \left(
\tau_{l_{2}}\right) \psi \left( \tau _{l_{1}}\right) e^{-S_{M}\left[ \psi
,\psi ^{\ast} \right] },  \label{eq:M-green-function}
\end{equation}%
with the partition function 
\begin{equation}
Z_{M} \equiv \int \mathcal{D}\left[ \psi ,\psi ^{\ast}\right] \ \text{e}%
^{-S_{M}\left[ \psi ,\psi ^{\ast} \right] }.
\label{eq:M-perturbed-partition-function}
\end{equation}

Since as $M$ tends to infinity, $S_{M}\left[ \psi ,\psi ^{\ast} \right] $
tends to $S\left[ \psi ,\psi ^{\ast } \right] $ with a convergence speed $%
\frac{1}{M}$, then $G_{M}\left( \tau_{l_{1}}, \tau _{l_{2}}\right)$ tends to 
$G\left( l_{1}\beta /M, \left( l_{2}+1\right) \beta /M\right)$ with the same
convergence speed. For this reason, one can approximate that in the
continuous time limit 
\begin{equation}
G\left(\frac{l_{1}}{M}\beta, \frac{l_{2}+1}{M}\beta \right) =
2G_{2M}\left(\tau_{2l_{1}}, \tau _{2l_{2}+1}\right) - G_{M}\left(\tau
_{l_{1}}, \tau _{l_{2}}\right).  \label{eq:extrapolation_of_green_function}
\end{equation}

Define the $M$-dependent particle density as 
\begin{equation}
\rho_{M}\left(\tau_{l}\right)\equiv G_{M}\left(\tau_{l}, \tau_{l}\right).
\label{eq:M-particle-density}
\end{equation}
As $M$ tends to infinity, it tends to the particle density in continuous
time limit. Then one can conclude that the $M$-dependent particle density $%
\rho_{M}$ tends to the particle density $\rho$ with a convergence speed $1/M$
as $M$ tends to infinity. Therefore, one can approximate that 
\begin{equation}
\rho \left(l\beta / M
\right)=2\rho_{2M}\left(\tau_{2l}\right)-\rho_{M}\left(\tau_{l}\right).
\label{eq:extrapolation_of_density}
\end{equation}
Eqs.(\ref{eq:extrapolation_of_green_function}, \ref%
{eq:extrapolation_of_density}) help to lower down the error of $%
O\left(1/M\right)$ caused by finite $M$ to $O\left(1/M^2\right)$.

\section{Algorithm\label{sec:algorithm}}

\subsection{Routine for the Green's function\label{sec:algorithm-lv1}}

The $HGW$ equations (\ref{eq:hgw_equations_in_fourier_space}) are
mathematically nonlinear equations of the Green's function $\tilde{G}$. To
solve the nonlinear equations, one can use the Broyden algorithm [\onlinecite%
{press_numerical_1992}]. The Broyden algorithm is designed to solve the
non-linear equations $F\left[X\right]=0$ with an initial value $X=X_{0}$.
This algorithm mainly contains two inputs, the nonlinear function $F$, and
the initial value $X_{0}$. In our cases, $X$ stands for the Green's function 
$\tilde{G}$, and $F$ stands for $\tilde{G}^{\prime}-\tilde{G},$ where $%
\tilde{G}^{\prime}$ is given by 
\begin{equation}
\tilde{G}^{\prime}=\frac{1}{\tilde{H}^{-1}+\mathcal{C}\left[\tilde{H},\frac{1%
}{\tilde{V}^{-1}-\mathcal{C}\left[\tilde{H},\tilde{G}\right]}\right]},
\label{eq:update_for_g}
\end{equation}
with the correlation functional 
\begin{equation}
\mathcal{C}\left[\tilde{X},\tilde{Y}\right]\left(\alpha \right)\equiv \frac{1%
}{\mathcal{N}}\sum_{\gamma}\tilde{X}\left(\alpha+\gamma \right)\tilde{Y}%
\left(\gamma \right),  \label{eq:notation_for_correlation}
\end{equation}
and $X_{0}$ stands for the initial value given by 
\begin{equation}
\tilde{G}_{0}\left(\alpha \right)=\frac{1}{T\left(\alpha \right)-\frac{U}{2}%
\rho_{0}-\Sigma_{0}\left(\alpha \right)},  \label{eq:initial_value_of_g}
\end{equation}
where the initial particle density $\rho_{0}\in \left(0,2\right)$ is given
randomly, and the initial self energy $\Sigma_{0}$ is also given randomly.
Note that the correlation (\ref{eq:notation_for_correlation}) can be fasten
by discrete Fourier transformation (DFT) algorithm [%
\onlinecite{press_numerical_1992}], and as a result, the complexity of one
iteration (\ref{eq:update_for_g}) is $\mathcal{O}\left(N\log N\right)$.

There might be multiple solutions to the nonlinear equations. In our
calculations, only one solution is found in the case that $U/t$ is
sufficiently small, or $\beta t$ is sufficiently small. However, multiple
solutions are found in the case of strongly coupling and low temperature.
Our strategy is setting gradients to $U$ or $\beta$, and then solving the
Green's function with different initial values for each parameter, and
finally choosing the solution continuous with $U$ or $\beta$.

To eliminate the error of $1/M$ of the Green's function in Matsubara time,
one can set different numbers of Matsubara time slices, and then make
extrapolation. In our calculations, $M$ is set to $512, 1024$ and $2048$. To
show $M$ is sufficiently large, one can verify $\left(2\rho_{2048} -
\rho_{1024}\right) - \left(2\rho_{1024} - \rho_{512}\right) $ is close to
zero. To obtain the density, one can use the approximation $\rho \doteq
2\rho_{2048} - \rho_{1024}$. To obtain the Green's function, one uses 
\begin{equation}
G\left(\frac{l_{1}}{1024}\beta, \frac{l_{2}+1}{1024}\beta \right) =
2G_{2048}\left(\tau_{2l_{1}}, \tau _{2l_{2}+1}\right) - G_{1024}\left(\tau
_{l_{1}}, \tau _{l_{2}}\right).
\label{eq:extrapolation-of-green-function-1024}
\end{equation}
Clearly, Green's function on only discrete Matsubara time can be obtained.
In addition, the particle density $n$ per site relates to $\rho$ through the
relation 
\begin{equation}
n \equiv n_{i}\left(\tau \right) = \rho_{i\uparrow}\left(\tau \right) +
\rho_{i\downarrow}\left(\tau \right).  \label{eq:particle-density}
\end{equation}

The numerical cost of the calculation of the Green function is analyzed as
follows.  For $U=2, T=0.125$ at half filling and $M=1024, N=16$, the typical
numerical cost is about $2.3$ seconds  running on a 32 - core CPU(2.6GHz). The
numerical cost is almost proportional to $MN^2$, and thus  is applicable to
complicated systems.

The parameters $U$ and $T$ influence the number of iterations, and then
influence the numerical cost. We set $M=1024$ and $N=16$. The numerical costs dependent on $U$ at $%
T=0.125$ are presented in Tab.\ref{tab:time_vs_u}, and the results demonstrate
that the numerical cost might be exponential in the Hubbard $U$. The numerical
costs dependent on $T$ at $U=2$ are presented in Tab.\ref{tab:time_vs_beta},
and the results demonstrate that the numerical cost is almost linear in $1/T$. 
Besides, for a good precision, one should increases $M$ as $U$ increases or $T$ decreases. According to our
experience, setting $M=[16\times U / T]$ yields a satisfactory precision
(after the extrapolation (\ref{eq:extrapolation_of_green_function})). With
these factors in consideration, the $HGW$ method should be applicable to the cases at sufficiently low temperature 
but not very large $U$.  

\begin{table}[h]
    \caption{Dependence of the numerical cost on the Hubbard $U$}
    \label{tab:time_vs_u}\centering
    \begin{tabular}{cccccc}
    \hline
    $U$ & 2.0 & 2.5 & 3.0 & 3.5 & 4.0 \\ \hline
    cost (seconds) & 2.257 & 5.919 & 7.743 & 16.429 & 38.945 \\ \hline
    \end{tabular}%
\end{table}

\begin{table}[h]
    \caption{Dependence of the numerical cost on the inverse temperature}
    \label{tab:time_vs_beta}\centering
    \begin{tabular}{ccccc}
    \hline
    $1/T$ & 8.0 & 16.0 & 24.0 & 32.0 \\ \hline
    cost (seconds) & 2.367 & 3.523 & 7.497 & 10.967 \\ \hline
    \end{tabular}%
\end{table}



\subsection{Routine for the density - density correlator\label%
{sec:algorithm-lv2}}

In the routine for the density - density correlator, there are mainly three
steps. First, calculate $\tilde{H},\tilde{W}$ for given $\tilde{G}$ and
parameters. Second, construct $\mathcal{M}$ and $b$, and solve the linear
equations (\ref{eq:lambda_equation_in_linear_form}) to obtain $\tilde{\Lambda%
}$. Third, calculate $\tilde{\chi}_{0}$ using Eq.(\ref%
{eq:chi-zero-def_fourier_result}), and calculate $\tilde{\chi}^{\text{cov}}$
using Eq.(\ref{eq:chi_cov_fourier_result}). The second step has the largest
complexity, up to $\mathcal{O}\left(N^{4}\right)$. 
The linear equations can be solved iteratively in a much faster speed than
the linear system solver. 

The charge correlator $\chi^{\text{ch}}$ relates to $\chi$ through the
relation 
\begin{align}
\chi^{\text{ch}}_{i_{1}, i_{2}}\left(\tau_{1}, \tau_{2}\right) & \equiv
\left \langle n_{i_{1}}\left(\tau_{1}\right) n_{i_{2}}\left(\tau_{2}\right)
\right \rangle - \left \langle n_{i_{1}}\left(\tau_{1}\right) \right \rangle
\left \langle n_{i_{2}}\left(\tau_{2}\right) \right \rangle  \notag \\
& = \sum_{\sigma_{1},\sigma_{2}} \chi_{i_{1}\sigma_{1},i_{2}\sigma_{2}}
\left(\tau_{1}, \tau_{2}\right) .  \label{eq:charge-correlator}
\end{align}
Note that 
\begin{equation}
\chi_{i_{1}\sigma_{1},i_{2}\sigma_{2}} \left(\tau_{1}, \tau_{2}\right) =
\left \langle \rho_{i_{1}\sigma_{1}}\left(\tau_{1}\right)
\rho_{i_{2}\sigma_{2}}\left(\tau_{2}\right) \right \rangle - \left \langle
\rho_{i_{1}\sigma_{1}}\left(\tau_{1}\right) \right \rangle \left \langle
\rho_{i_{2}\sigma_{2}}\left(\tau_{2}\right) \right \rangle.
\end{equation}

The charge susceptibility $\chi_{\text{c}} \equiv \partial n / \partial \mu $
in discrete time Matsubata action should satisfy 
\begin{equation}
\chi_{\text{c}} = \Delta \tau \sum_{l=0}^{M-1}\sum_{i} \left\langle
nn_{i}\left(\tau_{l}\right)\right\rangle_{\text{c}} = \Delta \tau \tilde{\chi%
}^{\text{ch}}\left(0, 0\right),  \label{eq: kappa-calculation}
\end{equation}
with the (discrete) Fourier transformation 
\begin{equation}
\chi^{\text{ch}}_{i_{1}i_{2}} \left(\tau_{1},\tau_{2}\right) = \frac{1}{MN^2}%
\sum_{k,m} \tilde{\chi}^{\text{ch}}\left(k, m\right) \text{e}^{\text{i}%
k\cdot \left( i_{1}-i_{2}\right) }\text{e}^{\text{i}\pi 2m\cdot \left( \tau
_{1}-\tau _{2}\right) }.  \label{eq: fft-chi-ch}
\end{equation}
The numerical cost of calculation of the two - body correlators is analyzed
as follows.  For $M=1024, N=4$, the typical numerical cost is about $3$
hours running on a 32 - core CPU(2.6GHz).  The numerical cost is almost
proportional to the square of $MN^2$, and is almost independent of $U$ and $T
$. The numerical cost is a bit large, and is not applicable to realistic
materials with the current algorithm. 

\section{$GW$ equations \label{sec:derivation-gw}}

The $GW$ approximation is based on Hedin's equations 
\begin{align}
G^{-1}(1,2) &= H^{-1}(1,2) - \Sigma(1,2),  \notag \\
\Sigma(1,2) &= - \int d(34)\ G(1,4) W(3,1) \Lambda(4,2;3),  \notag \\
W^{-1}\left(1,2\right) &= V^{-1}\left(1,2\right)-\Pi \left(1,2\right), 
\notag \\
\Pi \left(1,2\right) &= \int d(34)\
G\left(1,3\right)G\left(4,1\right)\Lambda \left(3,4;2\right),
\label{eq: hedin-equations}
\end{align}
with Hedin's vertex $\Lambda(1,2;3) = \delta G^{-1}(1,2) / \delta v(3)$.

One can make the simplest approximation for Hedin's vertex $\Lambda$, 
\begin{equation*}
\Lambda(1,2;3)\doteq \delta{H^{-1}(1,2)}/{\delta v(3)}=\delta(1,2)%
\delta(1,3),
\end{equation*}
to obtain the $GW$ equations 
\begin{align}
G^{-1}\left(1,2\right) & =H^{-1}\left(1,2\right)-\Sigma \left(1,2\right), 
\notag \\
\Sigma \left(1,2\right) & =-G\left(1,2\right)W\left(2,1\right),  \notag \\
W^{-1}\left(1,2\right) & =V^{-1}\left(1,2\right)-\Pi \left(1,2\right), 
\notag \\
\Pi \left(1,2\right) & = G\left(1,2\right)G\left(2,1\right).
\label{eq: gw_equations}
\end{align}

\bibliography{ref}

\end{document}